\documentclass[a4paper,11pt]{article}
\pdfoutput=1 % if your are submitting a pdflatex (i.e. if you have
\usepackage{jcappub} % for details on the use of the package, please
\usepackage[T1]{fontenc} % if needed

\usepackage{booktabs}
\usepackage{multirow}
\usepackage{dcolumn}
\usepackage{siunitx}
\usepackage{colortbl}
\usepackage{subfig}
\usepackage{array}
\usepackage[dvipsnames]{xcolor}

\newcommand\morehorsp{\rule[-2.25mm]{0mm}{7mm}}

\definecolor{darkred}{rgb}{0.7,0,0}
\definecolor{orange}{RGB}{75,0,130}

\newcommand{\ie}{\emph{i.e. }} 
\newcommand{\eg}{\emph{e.g. }}
\newcommand{\LCDM}{$\Lambda$CDM }

\newcommand{\bb}{\boldsymbol} 

\newcommand{\eref}[1]{Eq.(\ref{#1})}

\title{Strong Lensing Time Delay Constraints on Dark Energy: a Forecast.}

\author[a]{Banafshe Shiralilou}
\author[a,b]{Matteo Martinelli}
\author[a]{Georgios Papadomanolakis,}
\author[a]{Simone Peirone,}
\author[a,c]{Fabrizio Renzi,}
\author[a]{Alessandra Silvestri.}

\affiliation[a]{Institute Lorentz, Leiden University, PO Box 9506, Leiden 2300 RA, The Netherlands}
\affiliation[b]{Instituto de F\'isica T\'eorica UAM-CSIC, Campus de Cantoblanco, E-28049 Madrid, Spain}
\affiliation[c]{Physics Department and INFN, Universit\`a di Roma ``La Sapienza'', Piazzale Aldo Moro 2, 00185, Rome, Italy}

\emailAdd{shiralilou@lorentz.leidenuniv.nl}
\emailAdd{peirone@lorentz.leidenuniv.nl}
\emailAdd{papadomanolakis@lorentz.leidenuniv.nl}
\emailAdd{matteo.martinelli@uam.es}
\emailAdd{fabrizio.renzi@roma1.infn.it}
\emailAdd{silvestri@lorentz.leidenuniv.nl}

\abstract{
Measurements of time delays between multiple quasar images produced by strong lensing are reaching a sensitivity that makes them a promising cosmological probe. Future surveys will provide significantly more measurements, reaching unprecedented depth in redshift, making strong lensing time delay (SLTD) observations competitive with other background probes. 
We forecast constraints on the nature of dark energy from upcoming SLTD  surveys, simulating future catalogues with different numbers of lenses distributed up to redshift $z\sim 1$ and focusing on cosmological parameters such as the Hubble constant $H_0$ and parametrisations of the dark energy equation of state. 
We also explore the impact of our ability to precisely model the lens mass profile and its environment, on the forecasted constraints. We find that in the most optimistic cases, SLTD will constrain $H_0$ at the level of $\sim 0.1\%$, while the CPL equation of state parameters, $w_0$ and $w_a$, can be determined with errors $\sigma_{w_0}\sim 0.05$ and $\sigma_{w_a}\sim 0.3$, respectively. Furthermore, we investigate the bias introduced  
when a wrong cosmological model is assumed for the analysis. We find that the value of $H_0$ could be biased up to $10 \sigma$, assuming a perfect knowledge of the lens profile, when a $\Lambda$CDM model is used to analyse data that really belong to a $w$CDM cosmology with  $w=-0.9$. Based on these findings, we identify a consistency check of the assumed cosmological model in future SLTD surveys, by splitting the dataset in several redshift bins. Depending on the characteristics of the survey, this could provide a smoking gun for dark energy.}

\begin{document}
\maketitle
\flushbottom

\section{Introduction}
\label{sec:intro}
The phenomenon of cosmic acceleration, {\it i.e.} the late phase of accelerated expansion of the Universe, has posed a major challenge for Cosmology since it was first established in 1998~\citep{riess}. The standard cosmological model $\Lambda$CDM, with a cosmological constant $\Lambda$ as the candidate mechanism responsible for cosmic acceleration, has so far been the most successful model in describing both early Universe observations, such as Cosmic Microwave Background (CMB), as well as the late time dynamics of the Universe, probed by observations of Baryon Acoustic Oscillations (BAO), galaxy clustering and weak lensing.

Despite the successes of $\Lambda$CDM, recent observations highlighted a discrepancy between the value of the Hubble constant today, $H_{0}$, inferred from CMB observations and the local measurements performed through the distance ladder technique. While the former estimate of $H_0$ depends on the assumed cosmological model, the latter does not depend strongly on any cosmological assumption, as it relies on the observation of standard candles (type Ia supernovae) whose absolute luminosity is calibrated using Cepheids as an anchor. Recent estimates of $H_0$ obtained using the latter technique have been provided by the SH0eS team \cite{Riess:2016jrr}, with their latest value achieved exploiting observations of Cepheids in the Large Magellanic Cloud from the Hubble Space Telescope, $H_0=74.03\pm1.42$ km/s/Mpc \cite{Riess:2019cxk}.

The CMB estimates of $H_0$ rely instead on constraints of the size of the sound horizon at the last scattering surface ($\theta_*$) a measurement which allows to extrapolate bounds on the current expansion rate. This extrapolation however implies an assumption for the expansion history of the Universe. Assuming a $\Lambda$CDM background, measurements of the CMB from the Planck collaboration provide $H_0=67.36\pm0.54$ km/s/Mpc \cite{Aghanim:2018eyx}, a value which is in tension with the local measurement of the SH0eS collaboration at $4.4\sigma$.

There is currently no consensus on what is causing the discrepancy in the measure of the Hubble constant between low and high redshift data. One possibility is that the results are biased by neglected systematic effects on observational data (see e.g. \cite{Dhawan:2017ywl,Kenworthy:2019qwq,Rose:2019ncv,Colgain:2019pck,Martinelli:2019krf}), while, on the other hand, this tension could indicate that we need to abandon the $\Lambda$CDM assumption when extrapolating results to present time. Investigations of the latter possibility have highlighted how early time deviations from standard physics have the potential to ease the tension (see e.g. \cite{troubleH0,VM,CMBfluc,EDE,phenom,Lin:2019qug}), while other studies have tried to solve this issue allowing for non standard late time evolution, which might be produced by dynamical dark energy (DE) models, modified theories of gravity or interactions between DE and  dark matter (such as \cite{DiValentino:2017iww,Agrawal:2019dlm,Keeley:2019esp,Gerardi:2019obr,Martinelli:2019dau}).

In order to shed light on this tension, the value of $H_0$ has been determined also with other kind of observations. For example, the discovery of the first binary neutron stars merging event, GW170817 \cite{Abbott:2017xzu,TheLIGOScientific:2017qsa,Monitor:2017mdv,Coulter:2017wya} and the detection of an associated electromagnetic counterpart have lead to the  measurement $H_0 = 70^{+12}_{-8}$. Even though this constraint is much weaker than those obtained by SNe and CMB observations it is expected to significantly improve with the discovery of new merging events with an associated counterpart \cite{Hotokezaka:2018dfi,Chen:2017rfc}.

Along with standard sirens (as gravitational wave events are called nowadays because of their analogy with standard candles), observations of the time delay between multiply imaged strongly lensed system has become a compelling method to obtain measurements of $H_0$ together with other cosmological parameters. The observational method of SLTD was first proposed in 1964 and it can now produce precise, although cosmology dependent, estimations of the Hubble constant thanks to accurate measurements of the time delays between multiple images of specific lensed quasar \cite{holicowV}. The analysis of four well-measured systems from the H0LiCOW lensing program \cite{Suyu:2016qxx} has recently provided a bound on the Hubble constant of $H_0 = 72.5 \pm 2.1 $ assuming a flat $\Lambda$CDM cosmology \cite{Birrer:2018vtm}. While the H0LiCOW program is ambitiously aiming to bring the SLTD estimates of $H_0$ to the 1\% precision (see e.g. \cite{Suyu:2016qxx} and \cite{HolicowXIII} where a 2.4\% constraints on $ H_0 $ is obtained combining six well-measured lensing systems), observations of lensed system from future surveys, such as the Large Synoptic survey Telescope (LSST)  which should start taking data in 2023 \cite{LSSToverviewpaper}, are expected to significantly improve the number of well-measured strongly lensed systems \cite{LSST}. The increase in the number of observed lensed sources will also open the possibility to constrain non standard cosmologies, \eg extensions in the dark and neutrino sector (see \cite{HolicowXIII} for recent constraints on these extended parameter space from SLTD). SLTD is sensitive to the cosmological model through a combination of distances, but unlike Type Ia SNe, SLTD measurements do not require any anchoring to known absolute distances. Typically however, obtaining cosmological constraints with SLTD systems requires precise measurements and modeling of the mass profile and of the environment of the lens system in order to have systematics reasonably under control.
Future surveys, like LSST, are also expected to provide enough well-measured systems to allow sufficient statistics with a selected subset of lenses for which a precise modelling of the lens properties can be obtained. This will certainly limit the impact on the cosmological constraints of the uncertainties in the modeling of lens mass and environment. LSST, for instance,  has the advantage of having both the wide field-of-view to detect many quasars, and the frequent time sampling to monitor the lens systems for time delay measurements. Several thousand lensed quasar systems should be detectable with LSST, and, as shown in~\cite{oguriLSST},  around 400 of these should yield time delay measurements of high enough quality to obtain constraints on cosmological models~\cite{liaoLSST}.

It is timely to investigate the constraints on cosmological parameters that can be obtained from future observations of strongly lensed systems. 
In this paper we focus on simple extensions of the $\Lambda$CDM expansion history and forecast SLTD constraints on these, as well as on the current expansion rate $H_0$. We do so, by creating synthetic mock catalogues of future survey with variable number of lenses up to 1000 and building a Gaussian likelihood to compare data with theory. We also include estimates of the lens galaxy stellar velocity dispersion in our analysis. 

The paper is organized as follows. In Section~\ref{sec:tdtheo} we outline the connection between the time delays and the cosmological model, describe the theoretical modeling of the lens velocity dispersion and illustrate how it can improve the time delay constraints on DE parameters. In Section~\ref{sec:datalikelihood} we outline our analysis method, describing the likelihood expression used to infer the posterior distributions of cosmological parameters, and  explaining the procedure used to generate mock datasets, which are used to forecast the constraints displayed in Section~\ref{sec:results}. In Section~\ref{sec:FOM} we discuss the constraining power of SLTD on the DE models of interest, assessing at the same time how observational uncertainties on the lens model and on line of sight effects impact the Figure of Merit of future surveys. Section~\ref{sec:shift} contains our investigation of the possible bias brought on the inferred parameters by a wrong assumption of the underlying cosmological model. We also propose a consistency check that could be performed on future SLTD datasets to verify this possibility. Finally, we summarize our conclusions in Section~\ref{sec:conclusions}.

%%%%%%%%%%%%%%%%%%%%%%%%%%%%%%%%%%%%%%%%%%%%%%%%%%%%%%%%%%%%%%%%%%
\section{Cosmology with Time Delay Measurements}\label{sec:tdtheo}
We shall describe the connection between gravitational lensing time delay and the cosmological model, and how we account for the velocity dispersion of the lensing galaxy  for our cosmological inference. We also specify the lens and environment mass modeling used for our analysis and include a description of the mass-sheet degeneracy, which provides a transformation of the lens mass profile that has no
observable effect other than to rescale the time delays \citep{falco1985}.
%%%%%%%%%%%%%
\subsection{Theory of Gravitational Lensing Time Delays}
In strongly lensed systems, the time that light rays take to travel between the source and the observer depends sensibly both on their path and on the gravitational potential of the lens.
For a given\emph{i}-th light ray, the time delay with respect to its unperturbed path is given by \cite{Grav_lensing1992,Grav_lensing2006}): 
\begin{equation}\label{eq:timedelay}
    t(\boldsymbol{\theta_i,\beta})=(1+z_{l})\frac{D_{l}D_{s}}{c\,D_{ls}}\left[\frac{(\boldsymbol{\theta_i}-\boldsymbol{\beta})^{2}}{2}-\psi_{_\bot}(\boldsymbol{\theta}_i)\right]\,,
\end{equation}
%%%%%%SLTD geometry
\begin{figure}[ht!]
\centering	
\includegraphics[width=0.6\textwidth,keepaspectratio]{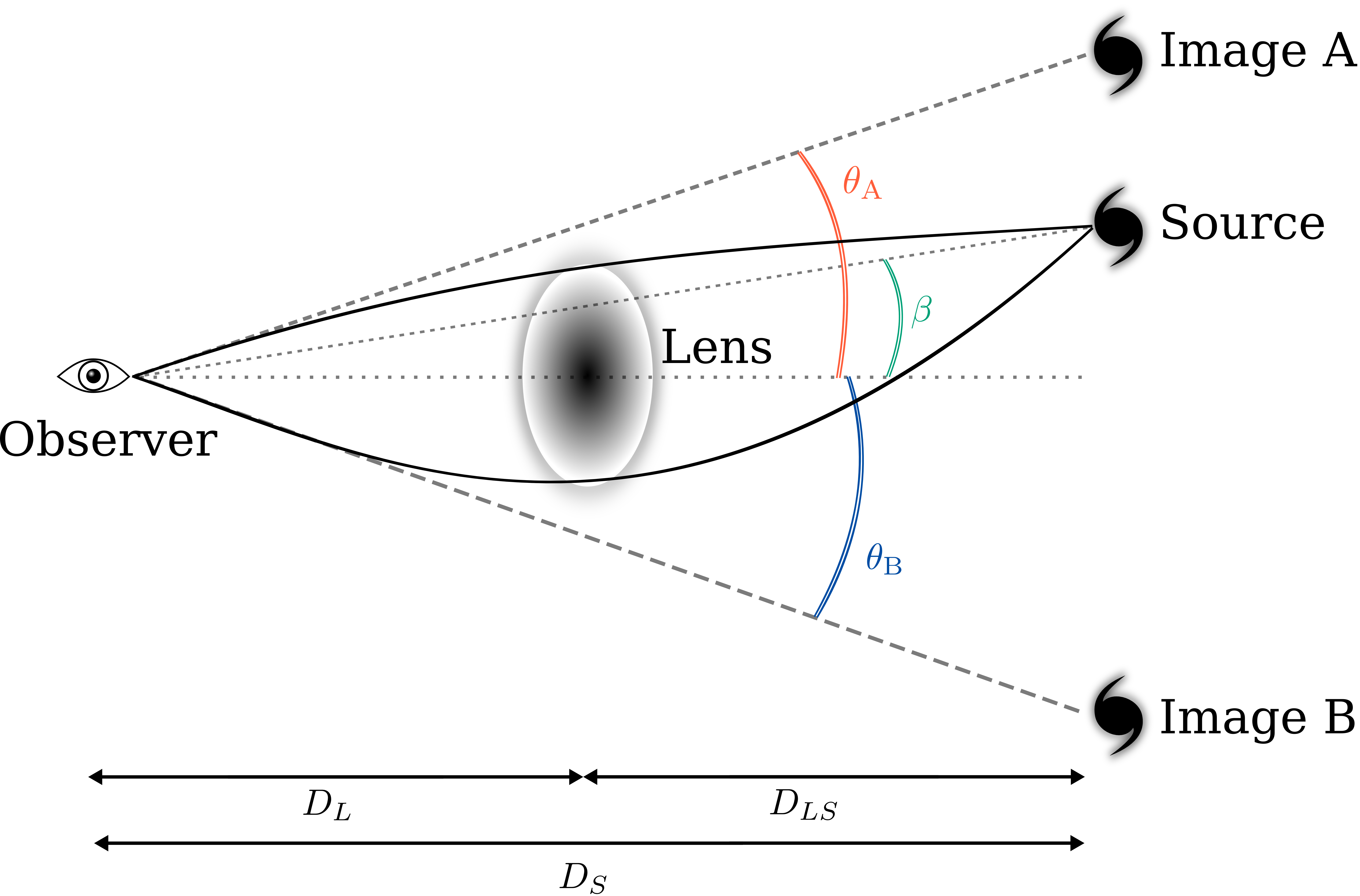}
\caption{The schematic view of a strongly lensed system. $D_{l}$, $D_{s}$ and $D_{ls}$ are, respectively, the angular diameter distance from the observer to the lens, from the observer to the source, and from the lens to the source. The solid angles $\beta$ and $\theta_i$ indicate the position of the source and the images, with respect to the lens plane.}
\label{fig:sltd}
\end{figure}\\
%%%%%%%%%%%%%%%%%%%%%%%%%%
where, as shown in Figure \ref{fig:sltd}, $\boldsymbol{\beta}$ and $\boldsymbol{\theta}_i$ stand, respectively, for the source and  the image position, $z_l$ is the redshift of the lens and  $\psi_{_\bot}(\boldsymbol{\theta_i})$ is the projected gravitational potential calculated on the lens plane. $D_{l}$, $D_{s}$ and $D_{ls}$ are, respectively, the angular diameter distance from the observer to the lens, from the observer to the source, and from the lens to the source; they satisfy the relation
\begin{equation}
D_{ls}=\frac{(1+z_{s})Ds-(1+z_{l})D_{l}}{1+z_{s}}\,,
\end{equation}
where $z_s$ is the redshift of the source.
The Fermat principle provides us with a lens equation for the relative angle between the true position of the source and each of the, possibly multiple, images:
\begin{equation}\label{lens_equation}
\boldsymbol{\theta_i}-\boldsymbol{\beta}= \nabla \psi_{_\bot}(\boldsymbol{\theta_i}),
\end{equation}
where $\nabla$ is the transverse gradient computed on the plane orthogonal to the direction of propagation of light.
It can be shown~\cite{Grav_lensing1992,Grav_lensing2006} that the combination $(\boldsymbol{\theta_i}-\boldsymbol{\beta})^{2}/2-\psi_{_\bot}(\boldsymbol{\theta_i})$ in Eq.~\eqref{eq:timedelay} is only dependent on the geometry and mass distribution of the deflectors; it is usually referred to as the \emph{Fermat potential} $\phi(\boldsymbol{\theta_i, \boldsymbol{\beta}})$.

As Eq.~\eqref{eq:timedelay} shows, the background cosmological parameters impact the gravitational lensing time delays through the ratios of angular diameter distances. In a flat Universe the angular diameter distance can be written as

\begin{equation}\label{eq:angdist}
    D(z)=\frac{c}{H_{0}(1+z)}\int^{z}_{0}\frac{dz'}
    {E(z')},
\end{equation}
where $E(z) = H(z)/H_0$ is the dimensionless Hubble rate, and $c$ is the speed of light. 

The relative time delay between two images A and B of a lensed system is given by the difference in the excess time of the two images, which can be rewritten in a simple form using the Fermat potential
\begin{equation}\label{eq:relativetimedelay}
    \Delta t_{AB}=(1+z_{l})\frac{D_{l}D_{s}}{c\,D_{ls}}\left[\phi(\boldsymbol{\theta_A, \boldsymbol{\beta}}) - \phi(\boldsymbol{\theta_B, \boldsymbol{\beta}})\right],
\end{equation}
where we can isolate the factor containing the dependence on cosmological parameters 
\begin{equation}\label{eq:Ddeltat}
    D_{\Delta t}= (1+z_l)\frac{D_{l}D_{s}}{c\,D_{ls}}\,,
\end{equation}
which is referred to as the \emph{time delay distance}.
\subsection{Lens Mass Model and Mass-Sheet degeneracy}
While assuming a cosmological model is enough to define $D_{\Delta t}$ through Eq. \eqref{eq:Ddeltat}, in order to be able to obtain theoretical predictions for $\Delta t_{AB}$, the Fermat potential needs to be computed. This requires a modeling of the mass profile of the lens galaxy. While an accurate description of the mass profile is challenging both experimentally and theoretically, the profiles of most discovered lens galaxies have been shown to be well fitted by a nearly elliptical power-law mass distribution~\citep{koopmans2006}. Throughout this paper,
when describing our lens systems, we will use a projected potential on the lens plane $\psi_{\perp}$ obtained assuming the Softened Power-law Elliptical Potential (SPEP)~\cite{Barkana_1998}:

\begin{equation}\label{eq:SPEP}
\psi_{\rm SPEP}(\boldsymbol{\theta})=
\frac{2A^2}{(3-\gamma')^2}\left[\frac{\theta_{1}^2 + \theta_{2}^{2}/q^2}{A^2}\right]^{(3-\gamma')/2}\,,
\end{equation}
where $q$ is the galaxy axis ratio, $ A= \theta_{E}/[ \sqrt{q} (\frac{3-\gamma'}{2})^ {1/(1-\gamma')}]$ is an overall normalization factor depending on the Einstein radius $\theta_{E}$ and $ \gamma' \approx 2 $ is the slope of the mass profile [which we define in Eq.~\eqref{eq.rholoc}].  $\theta_1$ and $\theta_2$ are the projections on the lens plane of the two dimensional image position $\boldsymbol{\theta}$.

Additionally, as common in the modeling of the mass profile of quadrupole lenses (see \eg \cite{Grav_lensing1992,Grav_lensing2006}), we include in our modeling of the lens mass profile a constant external shear yielding a potential in polar coordinates of the form:

\begin{equation}\label{shear}
\psi_p(\vartheta,\varphi) \equiv \frac{1}{2}\theta^2\gamma_{\rm ext}\cos2(\phi-\phi_{\rm ext}).
\end{equation}
where $ \gamma_{\rm ext} $ and $\phi_{ext}$ are the shear strength and angle. It is worth stressing that both $ \psi_p $ and $ \psi_{\rm SPEP} $ contribute to the projected potential $ \psi_\perp $. When using time delay measurements in cosmology, a complicating factor arises from the so-called \emph{mass-sheet degeneracy} (for a detailed discussion see \eg \cite{Grav_lensing1992,Grav_lensing2006}). In fact, a transformation of the lens convergence $ \kappa(\bb{\theta}) = \bb{\nabla}\psi_\perp/2 $ of the form:
\begin{equation}\label{eq.MSD}
\kappa'(\bb{\theta}) = \lambda\kappa(\bb{\theta}) + (1-\lambda)
\end{equation} 
will result in the same dimensionless observables, e.g. image positions and shapes, but will rescale the time delays by a factor $ \lambda $. 
The additional mass term can be due to perturbers that are very massive or close to the lens galaxy (which may need to be included explicitly into the mass model and affects stellar kinematics) or to the structures that lie along the LOS (see \eg \cite{Saha:2000kn,Wucknitz:2002fd,suyu2010,dispersionintegration}). Both effects can be summed up into a constant external convergence term, $ \kappa_{\rm ext} = 1-\lambda $, due to the mass sheet transformation described by \eref{eq.MSD}. The neat effect is a rescaling of the value of the observed time delay distance \cite{Keeton:2002ug,McCully:2013fga}:
\begin{equation}\label{eq:kext}
D'_{\Delta t} = \frac{D_{\Delta t}}{1 - \kappa_{\rm ext}}
\end{equation} 
This degeneracy between the external convergence and the mass normalization of the lens galaxy, if not resolved, can lead to a biased inference of the cosmological parameters \cite{seljak}.
Such an effect can be reduced by the combination of lensing data with stellar kinematics measurements, tracing the internal mass distribution of the lens galaxy \cite{Suyu:2013kha}.

%%%%%%%%%%%%%
\subsection{Stellar Dynamics Modeling}
In order to model the measurable stellar velocity dispersion $\sigma_v$ we need to model the 3D gravitational potential of the lens galaxy $\Phi$, in which stars are orbiting. This potential will have contributions from the mass distributions of both the lens and the nearby galaxies physically associated with the lens. To model the stellar velocity dispersion we follow the analysis of \cite{suyu2010,dispersionintegration}.
The overall mass density associated to $\Phi$ can be approximated as a spherically symmetric power law profile: 
\begin{equation}\label{eq.rholoc}
\rho_{local} = \rho_0 \left(\frac{r_0}{r}\right)^{\gamma'}
\end{equation}
the overall normalization $\rho_0 r_0^{\gamma'}$ can be determined quite well by lensing measurements, since it is a function of the lens profile characteristic only, and can be written as \cite{suyu2010}:
\begin{equation}
	\rho_\text{local} (r)=\pi^{-1/2}\,(\kappa_\text{ext} -1) \,\Sigma_\text{cr} \, \,R_\text{E}^{\gamma'-1} \frac{\Gamma(\gamma'/2)}{ \Gamma \left(\frac{\gamma' -3}{2}\right) }\,r^{-\gamma'}\,,
\end{equation}
where $R_\text{E}$ is the Einstein radius and $\Sigma_\text{cr}$ is the critical surface density.
As in \cite{suyu2010}, to calculate the LOS velocity dispersion we follow \cite{galactic_dynamics}. The three-dimensional radial velocity dispersion $\sigma_r$ is then found solving a spherical jeans equation:
\begin{equation}\label{sphericaljean}
\frac{\partial(\rho^* \sigma_r^2)}{\partial r}+ \frac{2\beta_{ani}(r)\rho^*\sigma_{r}^2}{r}+\rho^*\frac{\partial \Phi}{\partial r}=0\,,
\end{equation}
where $\beta_{ani} = r^2/(r^2 + r_{\rm ani}^2)$ is the anisotropy distribution of the stellar orbits in the lens galaxy and $\Phi$ is the galaxy gravitational potential associated to the overall density of Eq.(\ref{eq.rholoc}). For the modeling of the stellar distribution $\rho^*$, we have assumed the Hernquist profile~\cite{Hernquist}
\begin{equation}
\rho^*(r)= \frac{I_{0}a}{2\pi r (r+a)^3}\,,
\end{equation}
with $I_{0}$ being a normalization factor, $a= 0.551r_\text{eff}$ and $r_\text{eff}$ being the effective radius of the lensing galaxy. 
The luminosity-weighted velocity dispersion $\sigma_s$ is then given by :
\begin{equation}\label{sigmas}
I(R)\sigma_{s}^2=2\int_{R}^{\infty}\left(1-\beta_{ani}\left(\frac{R}{r}\right)^2\right)\frac{\rho^* \sigma_{r}^2 r\,dr}{\sqrt{r^2-R^2}}.
\end{equation}
Here $R$ is the projected radius and $I(R)$ is the projected Hernquist profile.
Finally, the luminosity-weighted LOS velocity dispersion within a measuring device aperture $\mathcal{A}$ is : 
\begin{equation}\label{sigmav}
(\sigma_{v})^2=\frac{\int_{\mathcal{A}}[I(R)\sigma_s^2*\mathcal{P}]R\,dR\,d\theta}{\int_{\mathcal{A}}[I(R)*\mathcal{P}]R\,dR\,d\theta}\,.
\end{equation}	
where $ *\mathcal{P} $ indicate convolution with the seeing (see also \cite{suyu2010,dispersionintegration}).
A prediction of the measurable velocity dispersion $\sigma_v$ is therefore obtained accounting for the observational characteristics of the survey, i.e. through the convolution, over $\mathcal{A}$, of the product $I(R)\sigma_{s}^2$ with the seeing $\mathcal{P}$.
Note that the cosmological dependence of $ \sigma_v $ is contained only in the combination $\Sigma_{cr}R_E^{\gamma' - 1}$, therefore separate $ \sigma_v $ as:
\begin{equation}
	(\sigma_{v})^2 = (1-\kappa_{ext})\frac{D_s}{D_{ls}}\ \mathcal{F}(\gamma',\theta_{E},\beta_{\rm ani},r_{\rm eff})
\end{equation}
where the terms $\mathcal{F}$ accounts for the computation of the integral in Eq.\eqref{sigmav} without the cosmological terms, $\theta_E$ is the angle associated with the Einstein radius, and all the cosmological information is contained in the ratio $ D_s/D_{ls}$. In this work, we follow the spectral rendering approach of \cite{dispersionintegration} to compute the luminosity-weighted LOS velocity dispersion from Eq.\eqref{sigmav}.

%%%%%%%%%%%%%%%%%%%%%%%%%%%%%%%%%%%%%%%%%%%%%%%%%%%%%%%%%%%%%%%%%%

\section{Analysis method and mock datasets}\label{sec:datalikelihood}

The final goal of this paper is to assess how well future surveys of strongly lensed systems will  constrain cosmological parameters, with a particular focus on simple extensions to the $\Lambda$CDM model. We do so by comparing the theoretical predictions of different cosmological models with forecasted datasets, based on mock catalogues. In practice we aim at calculating the {\it posterior distribution} $P(\vec{\pi}|\vec{d})$ for a set of cosmological parameters $\vec{\pi}$ given the set of (forecasted) data $\vec{d}$. Using the Bayes theorem, this can be written as 
\begin{equation}\label{eq:post_ideal}
 P(\vec{\pi}|\vec{d})\propto P(\vec{d}|\vec{\pi})P(\vec{\pi}),
\end{equation}
where $P(\vec{d}|\vec{\pi})$ is the {\it likelihood} of $\vec{d}$ given $\vec{\pi}$, and $P(\vec{\pi})$ is the prior distribution.

This expression for the posterior distribution does not include possible nuisance parameters which would account for uncertainties in the modeling of the lensed system, its environment and LOS effects. We will first generalize it to include these  parameters and then marginalize over them in order to obtain the final distribution only for the cosmological parameters. We  consider  the following nuisance parameters $\vec{\pi}_{\rm nuis}=(r_{\rm ani},\kappa_{\rm ext},\gamma')$. The final posterior can be obtained as~\cite{suyu2010}
\begin{equation}\label{eq:post_realistic}
 P(\vec{\pi}|\vec{d})\propto\int{dr_{\rm ani}d\kappa_{\rm ext}d\gamma'P(\vec{d}|\vec{\pi},\vec{\pi}_{\rm nuis})P(\vec{\pi})P(r_{\rm ani})P(\kappa_{\rm ext})P(\gamma')}\,,
\end{equation}
where $P(r_{\rm ani})$, $P(\kappa_{\rm ext})$ and $P(\gamma')$ are the prior distributions on each nuisance parameter. Notice that we do not include here as nuisance parameters, the other terms that enter in the lens model, e.g. the Einstein radius $\theta_E$ and the external shear $\gamma_{\rm ext}$.
In this paper we assume these to be perfectly known, since we are mainly interested in analyzing the degeneracy between the $H_0$ and the parameters of the lens modeling that are expected to have the most important impact on its estimation, \ie the slope of the power law profile and the external convergence (see \eg \cite{Schneider:2013sxa}). The degeneracy between $ H_0 $, $ \gamma' $ and $ \kappa_{\rm ext} $ may, in turn, also affect the constraints on DE equation of state through the well-known degeneracy between $ H_0 $ and DE parameters. We leave the study of the impact on our results of the inclusion of the whole parameter space of the lens model for a future work.
As discussed in Section~\ref{sec:tdtheo}, in order to break the mass-sheet degeneracy SLTD surveys combine measurements of the time delay between different images ($\Delta t$) and of the projected velocity dispersion within the lens ($\sigma_v$). The latter contains also a dependence on the cosmological parameters. Hence, our data vector will be therefore composed of this pair of measurements for each lensed system included in the dataset, with $\vec{d}=(\vec{\Delta t}, \vec{\sigma}_v)$. In order to constrain our cosmological models, these measurements need to be compared with the theoretical predictions $\vec{\Delta t}^{\rm th}$ and $\vec{\sigma}^{\rm th}$. Assuming that a Gaussian likelihood this can be written as
\begin{equation}\label{eq:likelihood}
P(\vec{d}|\vec{\pi}) =\exp{\left[-\frac{1}{2}\left(\sum_{i,j}{\frac{(\Delta t_{i,j}^{\rm th}(\vec{\pi}) - \Delta t_{i,j})^2}{\sigma^2_{\Delta t_{i,j}}}}+\sum_i{\frac{(\sigma_i^{\rm th}(\vec{\pi}) - \sigma_{v,i})^2}{\sigma^2_{\sigma_{v,i}}}}\right)\right]},
\end{equation}
where the index $i$ runs over all the lensed systems in the dataset, $j$ runs over the image pairs for each of the systems and we assume there is no correlation between the measurements of different systems. For a given set of cosmological ($\vec{\pi}$) and lens model ($\vec{\pi}_{\rm nuis}$) parameters, the theoretical predictions $\Delta t_{i,j}^{\rm th}$ and $\sigma_{v,i}^{\rm th}$ can be obtained from Eq.~\eqref{eq:timedelay} and Eq.~\eqref{sigmav} respectively. We compute the angular diameter distances that appear in these equations using \texttt{EFTCAMB} \cite{Hu:2013twa,Raveri:2014cka}, a public patch to \texttt{CAMB} \cite{CAMB1,CAMB2}.

With these predictions we can then reconstruct the posterior distribution $P(\vec{\pi}|\vec{d})$ sampling the parameter space and computing the likelihood of Eq.~\eqref{eq:likelihood} for each sampled point. The parameter space is sampled through the public Monte-Carlo Markov-Chain (MCMC) code \texttt{CosmoMC} \citep{cosmomc}, with the parameter vector $\vec{\pi}$ including the total matter density $\Omega_m$, the Hubble constant $H_0$ and  $w_0$ and $w_a$, which  parameterize the DE equation of state via the CPL form  \cite{cp,l}
\begin{equation}
 w(z) = w_0+w_a\frac{z}{1+z}.
\end{equation}
Using this parameterization, the dimensionless Hubble rate $E(z)$ appearing in Eq.~\eqref{eq:angdist} can be written as
\begin{equation}
 E(z) = \sqrt{\Omega_m(1+z)^3+\Omega_{\rm DE}(1+z)^{3(1+w_0+w_a)}\exp{\left[-w_a\frac{z}{1+z}\right]}}.
\end{equation}

In the following, we will explore three different DE models and this will determine whether or not we sample $w_0$ and $w_a$. The cases we investigate are:
\begin{itemize}
 \item $\Lambda$CDM, where both parameters are fixed to $w_0=-1$ and $w_a=0$, recovering the standard cosmological constant equation of state $w(z)=-1$;
 \item $w$CDM, where $w_a=0$, but we keep $w_0$ free to vary, obtaining a constant equation of state which might however deviate from $-1$;
 \item $w_0w_a$CDM, where both $w_0$ and $w_a$ are free to vary and we explore the possibility of a DE with a time dependent equation of state.
\end{itemize}
We always assume a flat Universe, with the DE density set by the relation $\Omega_{\rm DE}=1-\Omega_m$.

As stated above, sampling only over the parameters $\vec{\pi}$, while keeping the nuisance parameters $\vec{\pi}_{\rm nuis}$ fixed to their fiducial values, implicitly assumes that the lensed system is perfectly known: we label such cases as {\it ideal}. We consider also a {\it realistic} cases, where the nuisance parameters are allowed to vary. In Table \ref{tab:priors} we show the prior distributions assumed for all the parameters, with the cosmological ones always sampled using a uniform prior $[\pi^i_{\rm min},\pi^i_{\rm max}]$. In the $w$CDM and $w_0w_a$CDM cases, we additionally impose an {\it acceleration prior}, which limits the DE equation of state to $w(z)<-1/3$. In the realistic case, we additionally sample the nuisance parameters using Gaussian priors $\mathcal{G}$ for $\kappa_{\rm ext}$ and $\gamma'$ and a uniform prior for $r_{\rm ani}$. Keeping the nuisance parameters fixed, like in the ideal case, effectively amounts to using Dirac $\delta_D$ distributions as their priors. As stated by~\citep{suyu2010}, we stress that the uncertainty in $r_{\rm eff}$ has a negligible effect on the velocity dispersion modeling.

\begin{table}[!ht]
\begin{center}
    \setlength\extrarowheight{+6pt}
\begin{tabular}{ccc}
\toprule
    Parameter                    & Ideal case        & Realistic case \\[.7em]
\hline
\hline
$\Omega_m$                       & $[0,1]$           & $[0,1]$   \\
\morehorsp
$H_0$                            & $[40,140]$        & $[40,140]$   \\
\morehorsp
$w_0$                            & $[-3,0]$          & $[-3,0]$   \\
\morehorsp
$w_a$                            & $[-4,4]$          & $[-4,4]$   \\
\morehorsp
$\kappa_{\rm ext}$               & $\delta_D(-0.03)$ & $\mathcal{G}(-0.03,0.05)$\\
\morehorsp
$\gamma'$                        & $\delta_D(1.93)$  & $\mathcal{G}(1.93,0.02)$\\
\morehorsp
$r_{\rm ani}\ (\si{\arcsecond})$ & $\delta_D(3.5)$   & $[0.665,6.65]$\\
\bottomrule
\end{tabular}
\caption{Prior ranges on the cosmological and nuisance parameters sampled in our analysis.}\label{tab:priors}
\end{center}
\end{table}

\subsection{Mock Catalogues}\label{sec:mock}
The last ingredient that we need in order to compute the likelihood, is the data vector $\vec{d}$. We generate three mock catalogues containing different numbers of observed systems, i.e. with $N_{\rm lenses}=10,\ 100,\ 1000$ lenses, uniformly distributed in the redshift range $0<z\leq1$. Furthermore, we assume all the systems in the dataset to be identical to each other, adopting for all of them the mass profile described in Eq.~\eqref{eq:SPEP}, with the fiducial values of the model parameters set to those of the H0LiCOW resolved quadruply lensed system HE0435-1223 \cite{holicowIV}, listed in Table \ref{table:parameters}. We also assume that the redshift difference between the lens and the background source is the same for all the systems, with $\Delta z=1.239$.

\begin{table}[!ht]
\begin{center}
    \setlength\extrarowheight{+6pt}
\begin{tabular}{cccccccccc}
\toprule
 
Parameter & $\theta_{E}\ (\si{\arcsecond})$ & $q$  & $\theta_{q}\ (\si{\degree})$ & $\gamma'$ & $\gamma_{ext}$ & $\varphi_{\rm ext}\ (\si{\degree})$ & $\kappa_{\rm ext}$ & $r_{\rm eff}\ (\si{\arcsecond})$ & $r_{\rm ani}\ (\si{\arcsecond})$\\[.7em]
\hline
\hline
\morehorsp
Value     & $1.18$                         & $0.8$ & $-16.8$                      & $1.93$    & $0.03$         & $63.7$                              & $-0.03$            & $1.33$                           & $3.5$\\
\bottomrule
\end{tabular}
\caption{Fiducial values for the mock lens parameters.}\label{table:parameters}
\end{center}
\end{table}

Apart for being one of the currently best observed system, HE0435-1223 has the advantage of being almost symmetric in terms of the image configuration, which itself implies that the image separations are more than the observational threshold of $1''$ \cite{Oguri:2005je}. Moreover, the amount of external shear induced from the environment of this lens is consistent with the shear parameter of most of the studied lens systems in the assumed redshift range (see \eg the catalogue of \citep{oguriLSST}). Our assumptions on the redshift range of the dataset are justified by the constraints on source quasar redshifts from current surveys such as SDSS ~\cite{quasarredshift}, as well as the predicted peak lens and source redshift ranges for the LSST survey~\citep{LSST}.
Notice that our modeling of the mock catalogue implies also that we assume the same external convergence $\kappa_{\rm ext}$ for all systems. We fix the value of the external convergence on the best fit value of the distribution of $ \kappa_{\rm ext}$ estimated by the analysis of the environment of the lens HE0435-1223 \cite{Rusu:2006}. 
Since $\kappa_{\rm ext}$ is a LOS effect, this assumption might easily break down for surveys with an extended redshift range, for which a redshift dependence of this nuisance parameters could be included. 
The external convergence might also carry cosmological information, in particular if one wants to explore deviations from General Relativity; the different evolution of Large Scale Structures in modified gravity theories might indeed imprint characteristic features in the effect that these structures have on SLTD measurements, which can in principle be exploited to constrain departure from the standard General Relativity description \cite{seljak}.\\

\noindent
In addition to the lens parameters, in order to generate our mock datasets we also need to assume a fiducial cosmology. We choose two different fiducials, thus creating two classes of mock data: 
\begin{itemize}
  \item $\Lambda$-mock, where the DE equation of state parameter is constant in time and equal to $w(z)=-1$ (thus assuming $w_0=-1$ and $w_a=0$), and the cosmological parameters are chosen to be in agreement with the constraints obtained by the Planck collaboration \cite{plank2018}, i.e. $\Omega_m=0.295$, $H_0=67.3$ Km/s/Mpc.
  \item $w$-mock, which differs from the $\Lambda$-mock only in the value of the DE equation of state parameter, which is again constant but set to $ w(z)= -0.9$.
\end{itemize}
In both cases, we assume a flat Universe, with $\Omega_{\rm DE}=1-\Omega_m$.

\begin{figure}[t!]
\centering	
\includegraphics[width=0.6\textwidth,keepaspectratio]{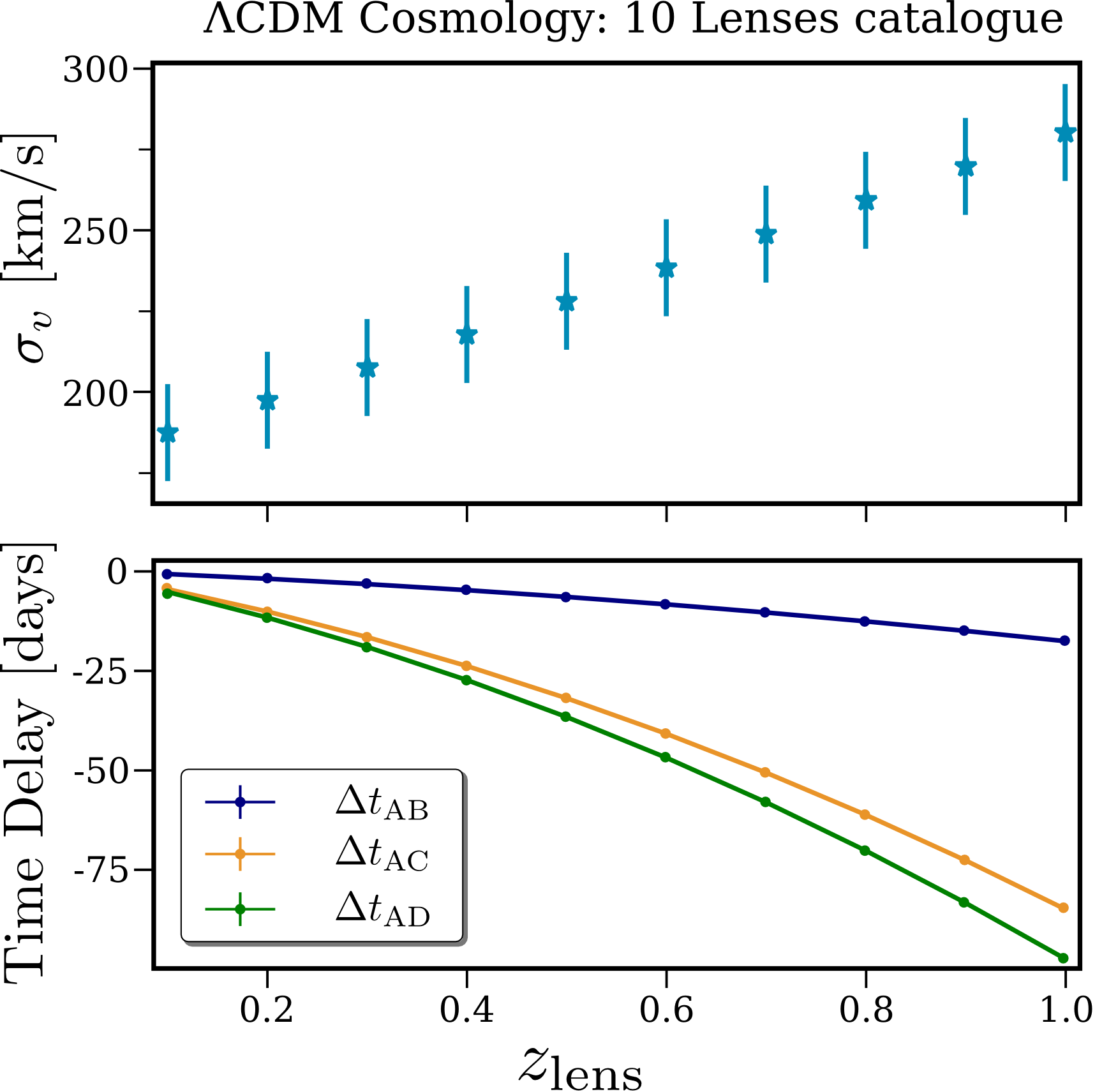}
\caption{\LCDM mock for a survey with $N_{\rm lens}=10$. The upper panel shows the velocity dispersion of the lens galaxies projected along the line of sight. 
The bottom panel shows the absolute time delay differences between image $A$ and the other three images.} 
\label{fig:mock}
\end{figure}

Once the lens and fiducial cosmological parameters are assumed, the relative time delays and the velocity dispersion can be computed following Eq.~\eqref{eq:relativetimedelay} and Eq.\eqref{sigmav}. Computing these for each of the $N_{\rm lens}$ lensed system contained in our dataset allows us to create our simulated data points; for each of these we assume that the time delays are observed with an error of $\sigma_{\Delta t}=0.8$ days\footnote{Based on our generated time delay values, this estimate fulfills the requirement of 0.2\% level time delay accuracy which, as pointed out by \citep{hojjati}, is needed for a low biased cosmological inference.}, while for the velocity dispersion measurements we assume a constant error $\sigma_{\sigma_v}=15\, \mathrm{km/s}$.
As an example, we show in Figure \ref{fig:mock} the $\Lambda$-mock obtained for a forecasted survey of $N_{\rm lens}=10$ lensed systems.

%%%%%%%%%%%%%%%%%%%%%%%%%%%%%%%%%%%%%%%%%%%%%%%%%

\section{Forecasts for cosmological parameters}\label{sec:results}

In this Section we present the forecasted bounds on cosmological parameters obtained following the analysis procedure and the mock datasets described in Section \ref{sec:datalikelihood}. We focus on the $\Lambda$-mocks, containing $N_{\rm lens}=10,\ 100,\ 1000$ lensed systems, and we analyse them, both in the ideal and realistic cases, using the three DE models we introduced: $\Lambda$CDM, $w$CDM and $w_0w_a$CDM.

\paragraph{$\Lambda$CDM -}
In a standard \LCDM scenario, we find that future strong lensing surveys will be able to constrain $H_0$ at the same level of Planck~\cite{plank2018}, $\sigma_{H_0}\sim 1\%$, already with $N_{\rm lens}=10$ in the ideal case; this result is consistent with what was found in~\cite{Jee_2016} for a catalogue of 55 lenses. Increasing the number of  systems to $N_{\rm lens}=100$, improves the bound on $H_0$ by a factor of $\sim3$, while with our most optimistic dataset ($N_{\rm lens}=1000$) we find that $H_0$ could be constrained with an error of $\sim0.1\%$. These results are shown in the left panel of Figure~\ref{fig:lambdamock_lambdafit} and in the $\Lambda$CDM entries of Table \ref{tab:resnonuis} shown in Appendix \ref{app:cosmo}.

In the realistic cases, where the nuisance parameters are let free to vary, the constraints on the Hubble rate are worsened by a factor of $\sim4$ for $N_{\rm lens}=10$. This worsening is mainly due to the strong degeneracy between $H_0$ and $\kappa_{\rm ext}$ described by Eq.~\eqref{eq:kext}, which is clearly visible in the right panel of Figure~\ref{fig:lambdamock_lambdafit}. Increasing the number of lenses  improves the bounds on both parameters, and we reach a $\sim2\%$ constraint on $H_0$ when $N_{\rm lens}=1000$. All the results for the realistic cases are shown in Table~\ref{tab:resreal} in Appendix \ref{app:cosmo}.

\begin{figure}[t!]
\begin{center}
\begin{tabular}{cc}
\includegraphics[width=0.85\columnwidth]{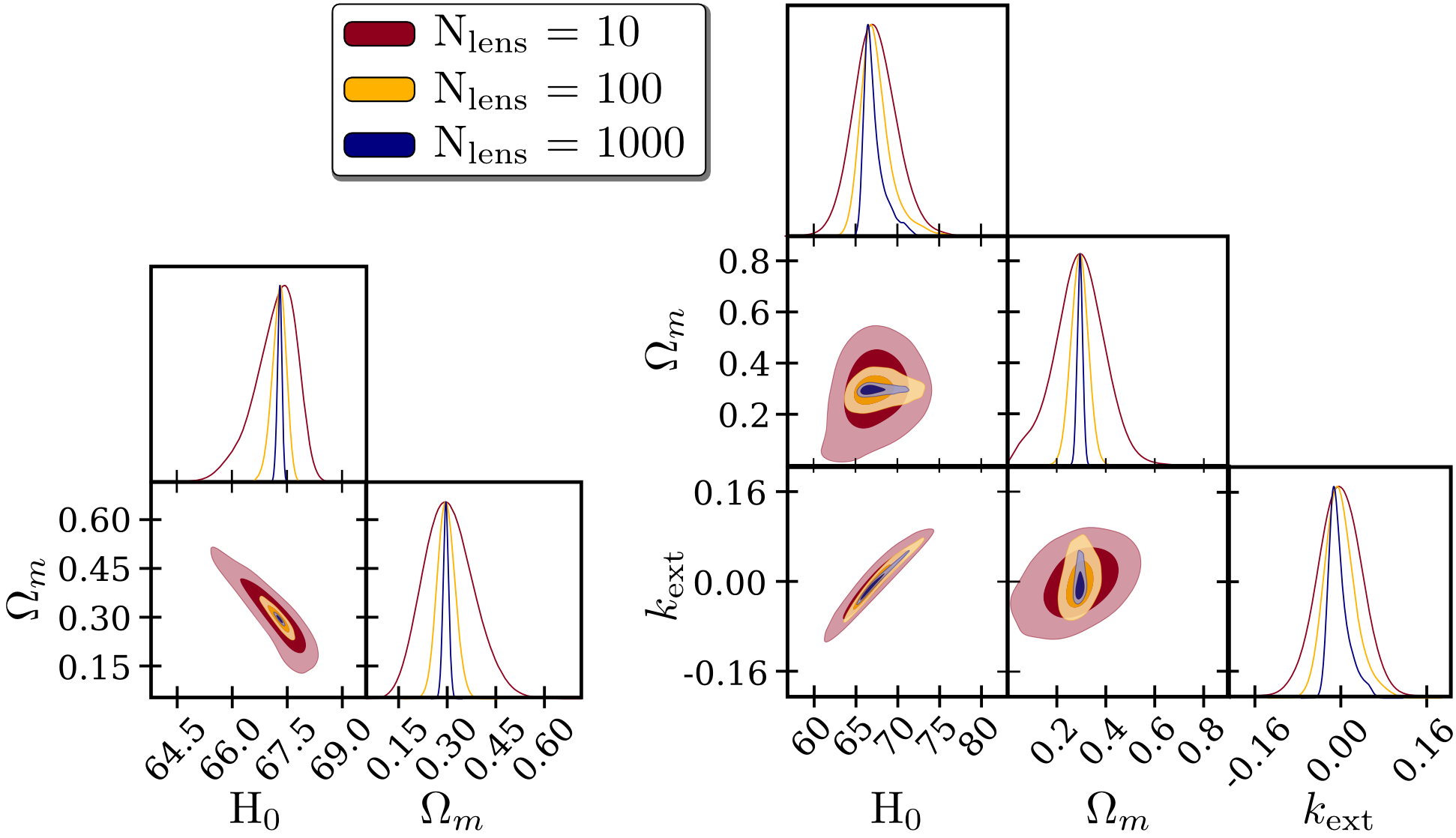}\\
\end{tabular}
\caption{Constraints on the $\Lambda$CDM cosmological model obtained using the $\Lambda$-mock datasets with $N_{\rm lens}=10$ (red contours), $100$ (yellow contours) and $1000$ (blue contours). The left panel shows the results for the ideal case, with nuisance parameters fixed to their fiducial values, while the right panel refers to the realistic case where these parameters are free. We show here only $\kappa_{\rm ext}$ as this is the only nuisance parameter with a significant degeneracy with the cosmological ones.} \label{fig:lambdamock_lambdafit}
\end{center}
\end{figure}

\paragraph{$w$CDM -}
Using the mock datasets to constrain this simple extended DE model, we find that in the ideal case with $N_{\rm lens}=10$, $H_0$ can now be measured with an error of $\sim 4\%$, which is improved to $\sim 1\%$ and $\sim 0.3\%$ for $N_{\rm lens}=100$ and $N_{\rm lens}=1000$ respectively. The parameter determining the equation of state for DE, $w_0$, is constrained at the level of $\sim 34\%$ for the $10$ lenses case, while moving to the optimistic $1000$ lenses configuration boosts the constraining power on this parameter up to $\sim 2\%$, thanks to the breaking of the degeneracy between $H_0$ and $w_0$. Such a result highlights how the improvement of SLTD measurements will significantly impact the investigation of DE alternatives to $\Lambda$CDM. These results are shown in Figure \ref{fig:lambdamock_wfit}, while the constraints on all the sampled parameters are included in Appendix \ref{app:cosmo} in Table \ref{tab:resnonuis}.

When considering the realistic case (see right panel of Figure \ref{fig:lambdamock_wfit} and Table \ref{tab:resreal}), the worsening of the constraints due to the nuisance parameters has a different trend with respect to the $\Lambda$CDM model; in the $10$ lenses case, the additional degeneracy introduced by $\kappa_{\rm ext}$ worsen the bounds on $H_0$ only by a factor $\sim 2$ (with respect to the factor $\sim 4$ of the $\Lambda$CDM case), due to the already existing degeneracy between $H_0$ and $w_0$, while in the $100$ and $1000$ lens cases, when this degeneracy is broken, the constraints become looser by a factor $\sim 5$ and $\sim 7$ respectively.

As $\kappa_{\rm ext}$ affects $w_0$ only through its degeneracy with $H_0$, moving from the ideal to the realistic case does not have an extreme impact on the DE parameter, with the constraints getting worse by a factor of $\sim 2 $ for $N_{\rm lens}=10,\ 100,\ 1000$ 

\begin{figure}[t!]
\begin{center}
\begin{tabular}{cc}

\includegraphics[width=0.99\columnwidth]{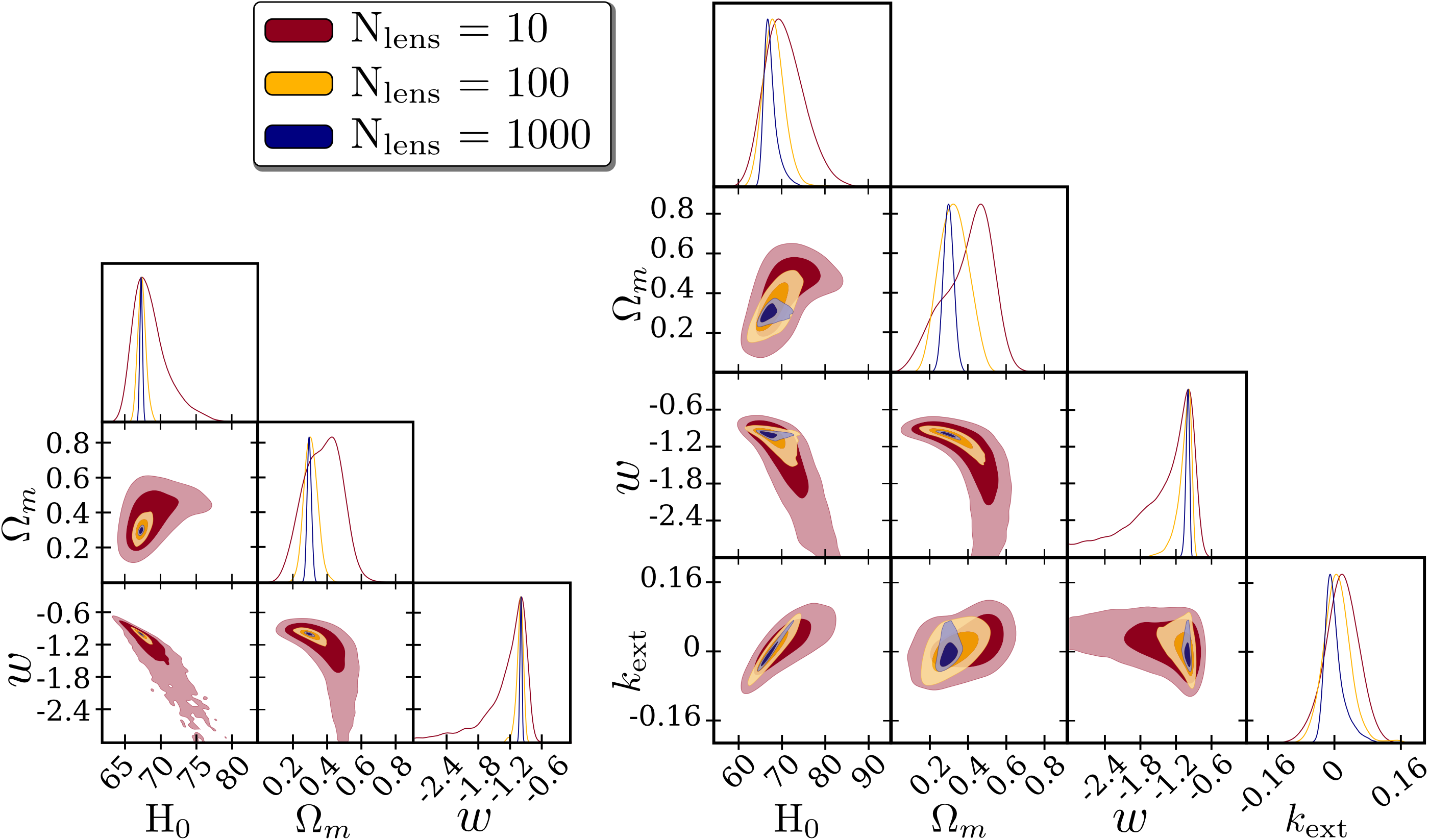}\\
\end{tabular}
\caption{Constraints on the $w$CDM cosmological model obtained using the $\Lambda$-mock datasets with $N_{\rm lens}=10$ (red contours), $100$ (yellow contours) and $1000$ (blue contours). The left panel shows the results for the ideal case, with nuisance parameters fixed to their fiducial values, while the right panel refers to the realistic case where these parameters are free. We show here only $\kappa_{\rm ext}$ as this is  the only nuisance parameter with a significant degeneracy with the cosmological ones.}
\label{fig:lambdamock_wfit}
\end{center}
\end{figure}

\begin{figure}[t!]
\begin{center}
\begin{tabular}{cc}
\includegraphics[width=\columnwidth]{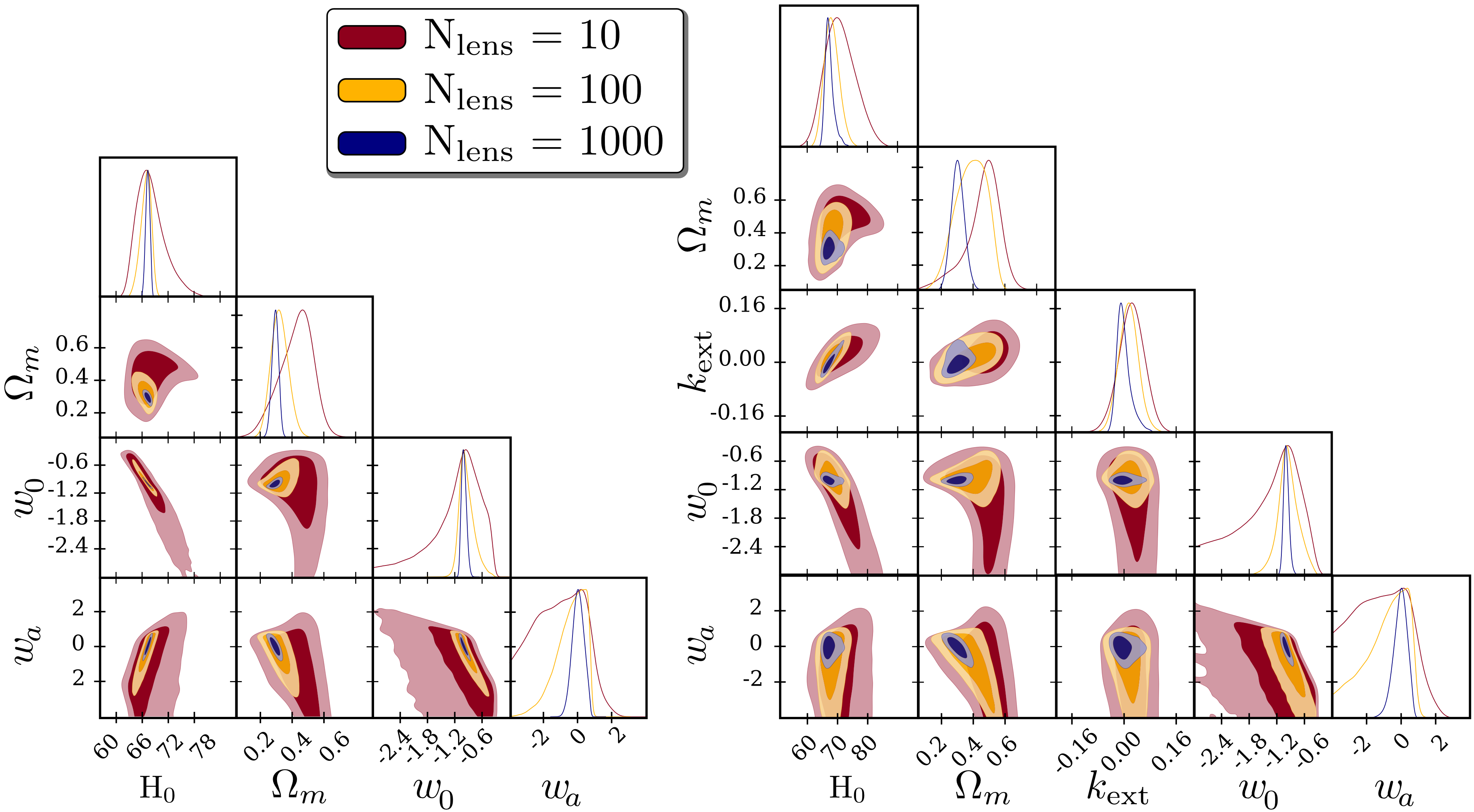}\\
\end{tabular}
\caption{Constraints on the $w_0w_a$CDM cosmological model obtained using the $\Lambda$-mock datasets with $N_{\rm lens}=10$ (red contours), $100$ (yellow contours) and $1000$ (blue contours). The left panel shows the results for the ideal case, with nuisance parameters fixed to their fiducial values, while the right panel refers to the realistic case where these parameters are free. We show here only $\kappa_{\rm ext}$ as this is the only nuisance parameter with a significant degeneracy with the cosmological ones.} \label{fig:lambdamock_w0wafit}
\end{center}
\end{figure}

\paragraph{$w_0w_a$CDM -}
In this case, we find that due to the degeneracies between $H_0$ and the DE parameters $w_0$ and $w_a$, the constraints on $H_0$ are significantly worsened. We see that a strong lensing survey could reach a $\sim 1\%$ level bound on the Hubble parameter only with the most optimistic configuration of this paper, i.e. $N_{\rm lens}=1000$ in the ideal case. Due to their degeneracy, $w_0$ and $w_a$  are not efficiently constrained solely with SLTD data; the best constraint is of the order  of $\sim 5 \%$ on $w_0$ and $\sigma_{w_a}\sim0.3$ on $w_a$, in the most optimistic case. Of course,  possible synergies of future SLTD surveys with other background probes, such as SNIa or BAO, would significantly improve this situation, breaking the degeneracy between the DE parameters and allowing to obtain again a bound on $H_0$ competitive with respect to CMB or local measurements.

The effect of nuisance parameters when considering the realistic case is similar to what is found for the $w$CDM case, with the additional parameters affecting mainly the bounds on $H_0$, whose error reaches now $\sim 2\%$ for $N_{\rm lens}=1000$, while not showing significant impact on the DE parameters.

The results for the $w_0w_a$ case are shown in Figure \ref{fig:lambdamock_w0wafit}, while numerical constraints are reported in Tables \ref{tab:resnonuis} and \ref{tab:resreal} shown in Appendix \ref{app:cosmo}.

\subsection{Figure of Merit for Strong Lensing Time Delay}\label{sec:FOM}
\begin{figure}[ht!]
\centering
\includegraphics[width=\textwidth,keepaspectratio]{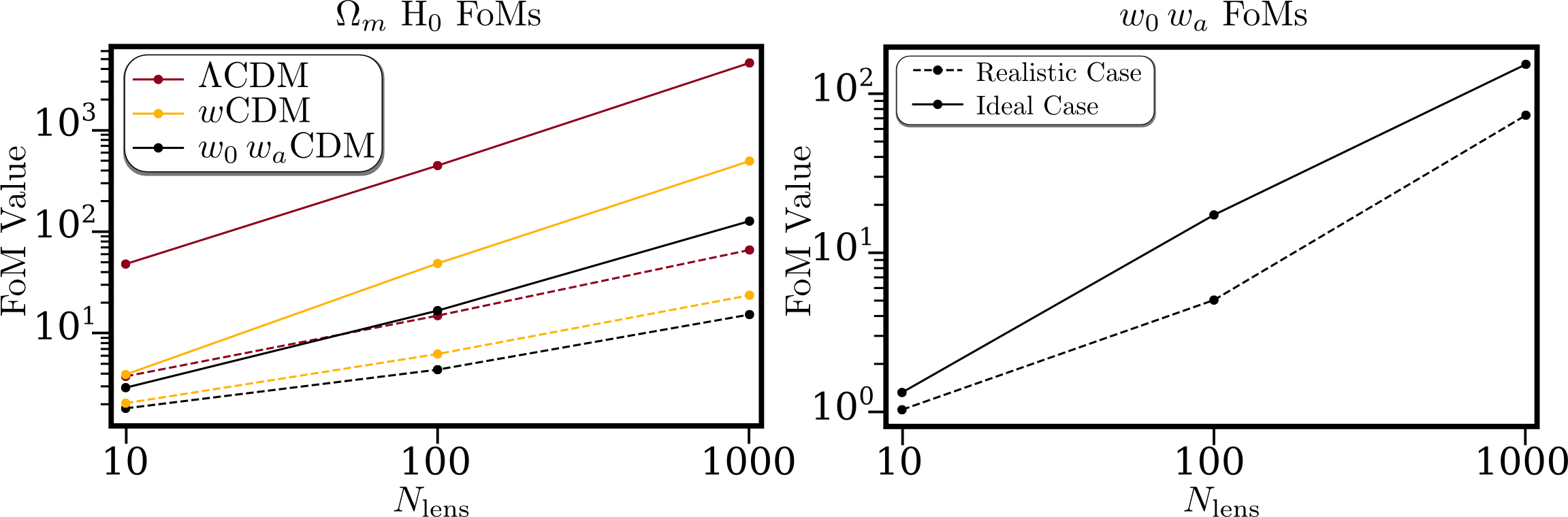}
\caption{FoM for the $\Omega_m,H_0$ (left panel) and $w_0,w_a$ (right panel) as a function of $N_{\rm lens}$ when analysing $\Lambda$CDM (red lines), $w$CDM (yellow lines) and $w_0w_a$CDM (black lines). The solid lines refer to the ideal case, while the dashed lines account for free nuisance parameters (realistic case).} \label{fig:FOM}
\end{figure}
We would like to quantify the constraining power of SLTD surveys, and its improvement with the number of observed systems, in a general way that allows to directly compare the performance of different surveys. For this purpose, we rely on the commonly used {\it Figure of Merit} (FoM) \cite{Albrecht:2006}. For two parameters $\alpha$ and $\beta$ the FoM is
\begin{equation}\label{eq:generalfom}
 {\rm FoM}_{\alpha\beta}=\sqrt{\det{\tilde{F}_{\alpha\beta}}},
\end{equation}
where ${\mathbf F}$ is the {\it Fisher information matrix} for a generic number of parameters and $\tilde{F}_{\alpha\beta}$ is the Fisher matrix marginalized over all the parameters except for $\alpha$ and $\beta$. Given its definition, the FoM gives an estimate of the area of the confidence contours for two parameters, thus quantifying the constraining power of an experiments on them, taking also into account their correlation. It is important to remember that  such a definition implies approximating the posterior distribution $P(\vec{\pi}|\vec{d})$ to a Gaussian.

From our MCMC analysis we derived a covariance matrix ${\mathbf C}={\mathbf F}^{-1}$, which contains all the sampled parameters. Let us focus on two cosmological parameters of interest that are common to all the models investigated in this work: $\Omega_m$ and $H_0$. We shall marginalize the covariance matrices over all the other parameters and then compare the constraining power of our mock datasets in each of the cases analysed, using the FoM for $\Omega_m$ and $H_0$:
\begin{equation}\label{eq:Homfom}
 {\rm FoM}_{\Omega_mH_0}=\sqrt{\det{\tilde{C}^{-1}_{\Omega_mH_0}}}.
\end{equation}
The posterior for $\Omega_m$ and $H_0$ is very close to a Gaussian one when the parameters are tightly constrained, e.g. in the $\Lambda$CDM case with $N_{\rm lens}=1000$; however, the gaussian approximation becomes less and less efficient as the number of lenses in the dataset decreases. Hence,  the FoM values for the less constraining cases might be overestimated.

In the left panel of Figure~\ref{fig:FOM} we show the trend of the FoM for the ideal (solid lines) and realistic (dashed lines) cases as a function of $N_{\rm lens}$. Comparing these two cases, we can notice how the improvement in constraining power brought by the number of lenses is less significant when the nuisance parameters are let free to vary. We also notice that the FoM for the ideal and realistic cases become more similar to each other as we go from $\Lambda$CDM to the more general DE parametrized by CPL. This is consistent with the trends that we have discussed in Section~\ref{sec:results}.

In the right panel of Figure \ref{fig:FOM} we also show the FoM for the $w_0$ and $w_a$ parameters, ${\rm FoM}_{w_0w_a}$. Such quantity is commonly used when quantifying the expected sensitivity of future experiment  to the DE sector. We find that SLTD surveys can reach values of ${\rm FoM}_{w_0w_a}\approx 100$ for $N_{\rm lens}=1000$ (in the ideal case), which is comparable with other future surveys, such as Euclid, expected to reach ${\rm FoM}_{w_0w_a}\approx 400$ with its primary probes, \cite{2011arXiv1110.3193L} or the ${\rm FoM}_{w_0w_a}\approx 100$ reached by the combination of Weak Lensing measurements from SKA1 and DES, together with Planck observations \cite{Bacon:2018dui}.

%----------------------------------
\section{A smoking gun for dark energy?}\label{sec:shift}

In Section~\ref{sec:results} we used the $\Lambda$-mock and constrained three DE models which contained the assumed fiducial cosmology as a limiting case. However, when real data will be available, we will have no a priori knowledge of the underlying cosmological model, and assumptions about the latter might affect the results. In this Section we test the impact of  wrong assumptions about the underlying cosmology on constraints from future surveys. To this extent, we consider the $w$-mocks, generated with a fiducial $w_0=-0.9$, and fit the data assuming instead a  $\Lambda$CDM cosmology. Given that the latter does not contain the true fiducial as a limiting case, we can quantify the sensitivity of future surveys on this assumption by computing the shift of the mean values obtained for cosmological parameters. In particular, for $H_0$ we have
\begin{equation}\label{eq:shift}
 S(H_0)=\frac{|H_0-H_0^{\rm fid}|}{\sigma_{H_0}},
\end{equation}
where the fiducial value is the one used to generate the mock data, i.e. $H_0^{\rm fid}=67.3$ km/s/Mpc, and we assume that the $H_0$ distributions obtained through our analysis can be approximated by a Gaussian of width $\sigma_{H_0}$. 

\begin{figure}[t!]
\begin{center}
\begin{tabular}{cc}
\includegraphics[width=.7\columnwidth]{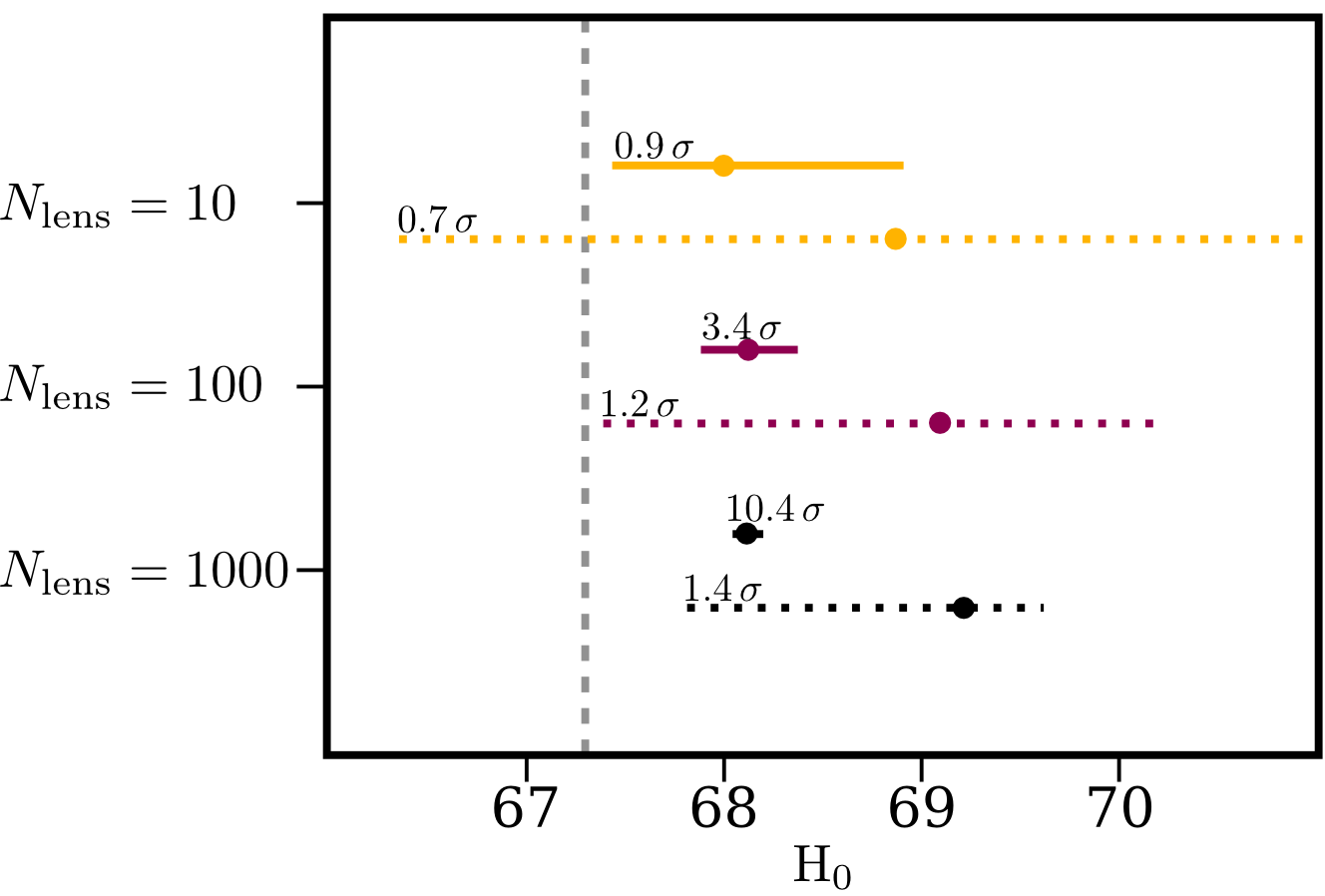} 
\end{tabular}
\caption{Marginalized means and error estimates on the value of the Hubble constant using the $w$-mocks ($w_0\neq -1$) analysed keeping the $w$ fixed to the $\Lambda$CDM value. The solid lines represent the ideal cases, while the dashed lines show the results of the realistic cases. The numbers above each line correspond to $S(H_0)$ computed following Eq. (\ref{eq:shift}).} \label{fig:H0shift}
\end{center}
\end{figure}

In Figure~\ref{fig:H0shift} we show the bounds on $H_0$ and the values of $S(H_0)$ changing the sample size, both for the ideal and realistic cases, when fitting the $w$-mocks with a $\Lambda$CDM cosmology. In the realistic case, the shift on this parameter is never statistically significant and reaches the maximum of $S(H_0)=1.4$ for the $1000$ lenses mock. However, in the ideal case the shift can be as high as  $10\sigma$. This implies that, if mass modelling of lenses reaches extreme accuracy with future surveys, the assumption of wrong cosmology could lead to significant tensions on $H_0$ value between SLTD observations and other independent cosmological measurements (e.g. from SH0eS \citep{Riess:2019cxk}). 

\begin{figure}[ht!]
\begin{center}
\begin{tabular}{cc}
\includegraphics[width=.70\columnwidth]{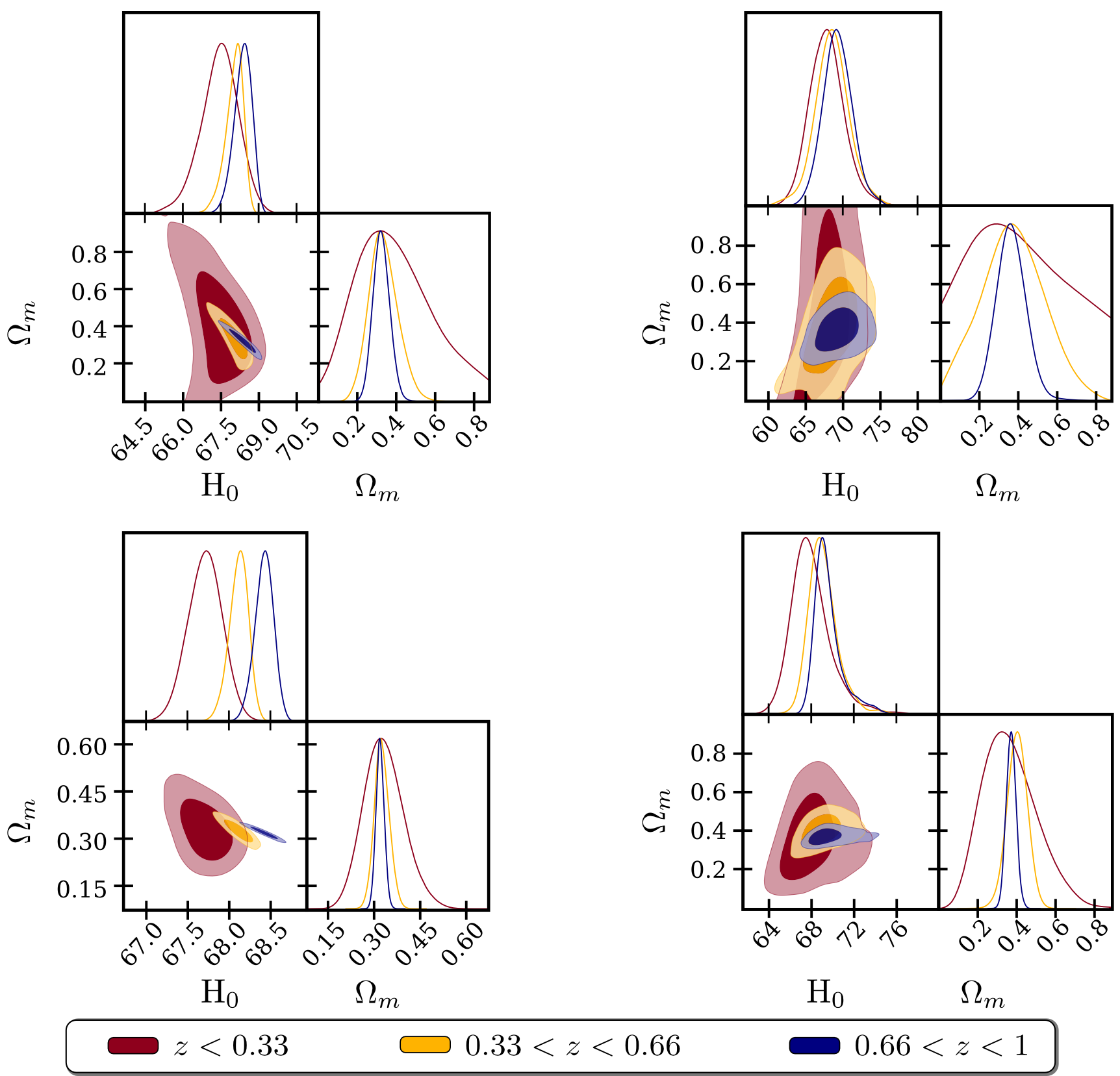} \\
\end{tabular}
\caption{Comparison of the constraints on $H_0$ and $\Omega_m$ from the analysis of the three datasets obtained splitting the original mock data in three redshift bins. Top panels refer to the $N_{\rm lens}=100$ dataset, while the bottom panels to the $N_{\rm lens}=1000$. Left panels do not include nuisance parameters (ideal case), while the right panels refer to the realistic case.} \label{fig:splitdata}
\end{center}
\end{figure}

Interestingly, it might be possible to exploit this shift effect,  to build a consistency check of the assumed cosmological model. Using a dataset of $N_{\rm lens}$ observed systems, we can split it in $N_{\rm bin}$ redshift bins and use the resulting datasets separately to constrain the parameters of a given cosmological model, e.g. $\Lambda$CDM. Should this model differ from  the ``true'' cosmology (or the fiducial one in the case of forecasts), the results obtained analysing separately the three datasets will be in tension with each other.
As a test case, we split the $w$-mock, both for $N_{\rm lens}=100$ and $N_{\rm lens}=1000$, in $N_{\rm bin}=3$ redshift bins and we fit these with a $\Lambda$CDM cosmology.  In Figure~\ref{fig:splitdata} we show the constraints on $H_0$ and $\Omega_m$ obtained through this analysis for $N_{\rm lens}=100$ (top panels) and $N_{\rm lens}=1000$ (bottom panels), with the left (right) panels showing the results in the ideal (realistic) case. While for $100$ overall lenses both the ideal and realistic case show no tensions on the cosmological parameters, in the ideal case with $N_{\rm lens}=1000$ a tension between the results on $H_0$ appears, with a $\sim 2\sigma$ significance between the first and the third bin. Such a result highlights how, with a sufficient number of observed systems, the assumption of a $\Lambda$CDM cosmology could be checked internally using only this observable; a statistically significant tension on the measured parameters in different redshift bins would  then  provide a smoking gun for the breakdown of $\Lambda$CDM, after internal systematics effect are excluded .

\section{Conclusions}\label{sec:conclusions}
In this paper we explored constraints on the nature of Dark Energy (DE) from future Strong Lensing Time Delay (SLTD) measurements. 
We simulated SLTD datasets starting from a fiducial cosmological model and  a description of the lens profile. For the latter, we assumed a common lens profile for all the systems. We distributed the lenses uniformly in the redshift range $0<z_{\rm lens}<1$, and we simulated the time delay that these would generate among different images of a background source, always placed at a $\Delta z=1.239$ from the lens, assuming different cosmologies.

In the ideal case, in which the lens profile and external environment parameters are perfectly known, SLTD measurements can provide constraints  that are competitive with other upcoming cosmological observations; $H_0$ can be constrained with an error as small as $\sim 0.1\%$ assuming a $\Lambda$CDM model and an optimistic dataset of $N_{\rm lens}=1000$ observed systems, while this error increases up to $\sim 1\%$ when the DE equation of state is allowed to vary.

We also evaluated the Figure of Merit (FoM) for the $w_0$ and $w_a$ in the Chevallier-Polarski-Linder parametrization of DE. We found that in our most optimistic case, the FoM can reach a value of $\sim 100$, which is competitive with what is expected from upcoming Large Scale Structure surveys. When considering a more realistic case, with the lens profile and lens environment parameters not perfectly known, the constraints worsen significantly. In particular $H_0$ is strongly degenerate with the nuisance parameter 
$\kappa_{\rm ext}$, that encodes the external convergence brought by additional structures along the line of sight between the observer and the lens. When we allow  $\kappa_{\rm ext}$ to vary (according to a prior), we find that $H_0$ can be constrained only up to $\sim 2\%$ both in the $\Lambda$CDM  and $w_0w_a$CDM cases. In the latter case, the FoM on $w_0,w_a$ can reach only $\sim 60$.

Furthermore we quantified the bias on cosmological parameters arising from a wrong assumption on the cosmological model in the analysis of future data. We analysed the mock dataset generated assuming $w=-0.9$ with a $\Lambda$CDM cosmology, i.e. with a fixed $w=-1$, and computed the shift $S(H_0)$ on the Hubble constant with respect to the fiducial value used to obtain the mock data. Interestingly, we found that in the ideal case this shift can reach $10\sigma$, highlighting how comparing the results obtained from SLTD observations with other measurements of $H_0$ could produce significant tensions on this parameter. Such a shift is however almost completely washed out in the realistic case, where $S(H_0)$ never exceeds $\sim 1.5\sigma$. 

The study of the shift in $H_0$, suggested an interesting, and potentially powerful, consistency check of the cosmological model, entirely based on SLTD data. We split our mock datasets constructed with a $w$CDM cosmology, with $w=-0.9$,  and analysed the three resulting datasets separately, (wrongly) assuming  $\Lambda$CDM cosmology. In the ideal case with $N_{\rm lens}=1000$, we found that the measurements of $H_0$ in the different bins would be in tension with each other up to $\sim 2\sigma$. This result shows how, with an accurate modeling of the observed lenses, future SLTD datasets can be used to internally test the assumptions on the cosmological model.

The future of SLTD looks bright; measurements are reaching the same accuracy of other, more traditional probes of background cosmology. As we have shown with our analysis, in the upcoming years, SLTD will provide competitive and complementary constraints on dark energy. 
It would be of great interest to not only further explore SLTD in the context of extended theories of gravity~\cite{MGtest1,MGtest2,MGtest3}, but also in combination with other cosmological probes. This work represents a first step in all these directions. In particular, the likelihood pipeline that we have built in CosmoMC will be of great use to explore complementarity of SLTD with other cosmological probes.

\acknowledgments
We thank Vivien Bonvin for helping with the H0LiCOW likelihood and dataset, Alessandro Sonnenfeld, Sherry Suyu and Isaac Tutusaus for useful comments and discussions. MM has received financial support through the Postdoctoral Junior Leader Fellowship Programme from la Caixa Banking Foundation (grant n. LCF/BQ/PI19/11690015). MM also acknowledges support from the D-ITP consortium, a program of the NWO that is funded by the OCW. AS and SP acknowledge support from the NWO and the Dutch Ministry of Education, Culture and Science (OCW), and from the D-ITP consortium, a program of the NWO that is funded by the OCW. FR  is supported by TASP, iniziativa specifica INFN. FR also acknowledges a visitor grant from the D-ITP consortium.

\clearpage
\appendix

\section{Constraints on cosmological parameters}\label{app:cosmo}

In this Appendix we show the constraints obtained on all free cosmological parameters when the $\Lambda$-mock is analysed. In Table \ref{tab:resnonuis} we show the results obtained in the ideal case, when the nuisance parameters are assumed to be perfectly known, while Table \ref{tab:resreal} shows the constraints in the realistic case, where also the $\kappa_{ext}$, $r_{ani}$ and $\gamma'$ are free. In these tables, for each parameter, we show the results obtained assuming different DE models, i.e. $\Lambda$CDM, $w$CDM and $w_0w_a$CDM.

%#############################table2
\begin{table}[!ht]
\begin{center}
    \setlength\extrarowheight{+6pt}
\begin{tabular}{ccccc}
\toprule 
    Parameter  & DE model    & 10 lenses                      & 100 lenses                   & 1000 lenses \\[0.7em]
\hline
\hline
 & $\Lambda$CDM     & $0.304^{+0.069}_{-0.085}$      & $0.296\pm 0.027$             & $0.2949\pm 0.0086$   \\
\morehorsp
$\Omega_m$ & $w$CDM & $0.37^{+0.12}_{-0.10}$         & $0.305^{+0.037}_{-0.042}$    & $0.296\pm 0.012$   \\
\morehorsp
 & $w_0w_a$CDM      & $0.425^{+0.12}_{-0.083}$       & $0.322\pm 0.053$    & $0.294^{+0.022}_{-0.020}$   \\
\hline\hline
 & $\Lambda$CDM     & $67.16^{+0.70}_{-0.41}$        & $67.28^{+0.21}_{-0.17}$      & $67.299^{+0.063}_{-0.057}$   \\
\morehorsp
$H_0$ & $w$CDM      & $68.1^{+2.1}_{-4.0}$           & $67.42^{+0.51}_{-0.62}$      & $67.31\pm 0.18$   \\
\morehorsp
 & $w_0w_a$CDM      & $68.3^{+3.0}_{-3.8}$           & $66.9^{+1.4}_{-1.2}$        & $67.33^{+0.52}_{-0.47}$   \\
\hline\hline
 & $\Lambda$CDM     & $-$                            & $-$                          & $-$   \\
$w_0$ & $w$CDM      & $-1.30^{+0.47}_{-0.10}$        & $-1.021^{+0.073}_{-0.046}$   & $-1.002^{+0.020}_{-0.018}$   \\
\morehorsp
 & $w_0w_a$CDM      & $-1.19^{+0.74}_{-0.21}$        & $-0.94^{+0.15}_{-0.19}$      & $-1.001^{+0.045}_{-0.063}$   \\
\hline\hline
 & $\Lambda$CDM     & $-$                            & $-$                          & $-$   \\
$w_a$ & $w$CDM      & $-$                            & $-$                          & $-$   \\
 & $w_0w_a$CDM      & $-1.1^{+1.8}_{-1.3}$           & $-0.43^{+1.0}_{-0.79}$        & $0.02\pm 0.34$   \\
\bottomrule
\end{tabular}
\caption{Mean marginalized values and their $68\%$ confidence level bounds for the three DE model considered. We show here the results for the ideal case for $10$, $100$ and $1000$ lenses.}\label{tab:resnonuis}
\end{center}
\end{table}
%#################################

\newpage

%#############################table
\begin{table}[!htp]
\begin{center}
      \setlength\extrarowheight{+6pt}
\begin{tabular}{ccccc}
\toprule
    Parameter  & DE model    & 10 lenses & 100 lenses & 1000 lenses \\[.7em]
\hline
\hline

 & $ \Lambda $CDM    & $0.292^{+0.11}_{-0.096}$      & $0.293\pm 0.035$        & $0.295\pm 0.011$   \\ 
\morehorsp
$\Omega_m$ & $w$CDM & $0.401^{+0.15}_{-0.091}$      & $0.327\pm 0.075$        & $0.299\pm 0.027$   \\
\morehorsp
 & $w_0w_a$CDM      & $0.460^{+0.13}_{-0.065}$      & $0.388^{+0.11}_{-0.084}$        & $0.304\pm 0.042$   \\

\hline\hline 

 & $ \Lambda $CDM    & $67.3^{+2.3}_{-2.7}$      & $67.5^{+1.1}_{-2.1}$        & $67.26^{+0.43}_{-1.5}$   \\
\morehorsp
$H_0$ & $w$CDM     & $70.8^{+3.6}_{-5.4}$      & $68.3^{+2.1}_{-2.8}$        & $ 67.51^{+0.77}_{-1.9} $   \\
\morehorsp
 & $w_0w_a$CDM     & $71.1^{+4.4}_{-5.8}$      & $68.2^{+2.2}_{-2.8}$        & $67.48^{+0.82}_{-1.8}$   \\
\hline
\hline

 & \textit{$\Lambda$}CDM     & $-$      & $-$        & $-$   \\
 \morehorsp
$w_0$ & $w$CDM      & $-1.44^{+0.62}_{-0.20}$      & $-1.07^{+0.15}_{-0.050}$        & $ -1.007^{+0.036}_{-0.024} $   \\
\morehorsp
 & $w_0w_a$CDM      & $-1.41^{+0.89}_{-0.37}$      & $-0.96\pm 0.23$        & $-0.997^{+0.050}_{-0.064}$   \\
\hline\hline
 & $ \Lambda $CDM     & $-$      & $-$        & $-$   \\
 \morehorsp
$w_a$ & $w$CDM      & $-$      & $-$        & $-$   \\
\morehorsp
 & $w_0w_a$CDM      & $-1.2^{+1.9}_{-1.6}$      & $-0.93^{+1.7}_{-0.64}$        & $-0.07^{+0.46}_{-0.33}$   \\
  
\hline\hline

 & $ \Lambda $CDM    & $-0.001\pm 0.041$      & $-0.001^{+0.023}_{-0.032}$        & $-0.0040^{+0.0092}_{-0.023}$   \\
 \morehorsp
$\kappa_{ext}$ & $w$CDM & $0.017\pm 0.041$      & $0.006\pm 0.034$        & $ -0.001^{+0.011}_{-0.027} $   \\
\morehorsp
 & $w_0w_a$CDM     & $0.023\pm 0.040$      & $0.011\pm 0.030$        & $-0.001^{+0.013}_{-0.025}$   \\

\hline\hline

 & $ \Lambda $CDM    & $> 3.06$      & $> 3.45$        & $4.1\pm 1.5$   \\
 \morehorsp
$r_{ani}$ & $w$CDM  & $> 3.46$      & $> 3.33$        & $  4.2\pm 1.4 $   \\
\morehorsp
 & $w_0w_a$CDM      & $> 3.57$      & $> 3.55$        & $4.2^{+1.9}_{-1.3}$   \\

\hline\hline

 & $ \Lambda $CDM    & $1.930\pm 0.018$      & $1.930\pm 0.013$        & $1.9301\pm 0.0052$   \\
 \morehorsp
$\gamma'$ & $w$CDM  & $1.927\pm 0.018$      & $1.930\pm 0.012$        & $ 1.9301\pm 0.0052 $   \\
\morehorsp
 & $w_0w_a$CDM      & $1.926\pm 0.018$      & $1.929\pm 0.013$        & $1.9300\pm 0.0052$   \\
\bottomrule
\end{tabular}
\caption{Mean marginalized values and their $68\%$ confidence level bounds for the three DE model considered. We show here the results for the realistic case for $10$, $100$ and $1000$ lenses.}\label{tab:resreal}
\end{center}
\end{table}
%#######################################

\bibliographystyle{unsrt}

\bibliography{biblio}

\begin{thebibliography}{10}

\bibitem{riess}
Peter Challis~\textit{et al.} Adam G.~Riess, Alexei V.~Filippenko.
\newblock 116(3):1009--1038, sep 1998.

\bibitem{Riess:2016jrr}
Adam~G. Riess et~al.
\newblock {A 2.4\% Determination of the Local Value of the Hubble Constant}.
\newblock {\em Astrophys. J.}, 826(1):56, 2016.

\bibitem{Riess:2019cxk}
Adam~G. Riess, Stefano Casertano, and Wenlong~\textit{et al.} Yuan.

\bibitem{Dhawan:2017ywl}
Suhail Dhawan, Saurabh~W. Jha, and Bruno Leibundgut.
\newblock {Measuring the Hubble constant with Type Ia supernovae as
  near-infrared standard candles}.
\newblock {\em Astron. Astrophys.}, 609:A72, 2018.

\bibitem{Kenworthy:2019qwq}
W.~D'Arcy Kenworthy, Dan Scolnic, and Adam Riess.
\newblock {The Local Perspective on the Hubble Tension: Local Structure Does
  Not Impact Measurement of the Hubble Constant}.
\newblock {\em Astrophys. J.}, 875(2):145, 2019.

\bibitem{Rose:2019ncv}
B.~M. Rose, P.~M. Garnavich, and M.~A. Berg.
\newblock {Think Global, Act Local: The Influence of Environment Age and Host
  Mass on Type Ia Supernova Light Curves}.
\newblock {\em Astrophys. J.}, 874(1):32, 2019.

\bibitem{Colgain:2019pck}
Eoin~\'O". Colg\'aiin.
\newblock {Recasting $H_0$ tension as $\Omega_m$ tension at low $z$}.
\newblock 2019.

\bibitem{Martinelli:2019krf}
Matteo Martinelli and Isaac Tutusaus.
\newblock {CMB tensions with low-redshift $H_0$ and $S_8$ measurements: impact
  of a redshift-dependent type-Ia supernovae intrinsic luminosity}.
\newblock 2019.

\bibitem{troubleH0}
Jos{\'e}~Luis {Bernal}, Licia {Verde}, and Adam~G. {Riess}.
\newblock {The trouble with H$_{0}$}.
\newblock {\em \jcap}, 2016(10):019, Oct 2016.

\bibitem{VM}
Eleonora Di~Valentino, Eric~V. Linder, and Alessandro Melchiorri.
\newblock Vacuum phase transition solves the $h_0$ tension.
\newblock {\em Phys. Rev. D}, 97:043528, Feb 2018.

\bibitem{CMBfluc}
Saroj {Adhikari} and Dragan {Huterer}.
\newblock {Super-CMB fluctuations can resolve the Hubble tension}.
\newblock {\em arXiv e-prints}, page arXiv:1905.02278, May 2019.

\bibitem{EDE}
Vivian {Poulin}, Tristan~L. {Smith}, Tanvi {Karwal}, and Marc {Kamionkowski}.
\newblock {Early Dark Energy can Resolve the Hubble Tension}.
\newblock {\em \prl}, 122(22):221301, Jun 2019.

\bibitem{phenom}
Prateek {Agrawal}, Francis-Yan {Cyr-Racine}, David {Pinner}, and Lisa
  {Randall}.
\newblock {Rock 'n' Roll Solutions to the Hubble Tension}.
\newblock {\em arXiv e-prints}, page arXiv:1904.01016, Apr 2019.

\bibitem{Lin:2019qug}
Meng-Xiang Lin, Giampaolo Benevento, Wayne Hu, and Marco Raveri.
\newblock {Acoustic Dark Energy: Potential Conversion of the Hubble Tension}.
\newblock 2019.

\bibitem{DiValentino:2017iww}
Eleonora Di~Valentino, Alessandro Melchiorri, and Olga Mena.
\newblock {Can interacting dark energy solve the $H_0$ tension?}
\newblock {\em Phys. Rev.}, D96(4):043503, 2017.

\bibitem{Agrawal:2019dlm}
Prateek Agrawal, Georges Obied, and Cumrun Vafa.
\newblock {$H_0$ Tension, Swampland Conjectures and the Epoch of Fading Dark
  Matter}.
\newblock 2019.

\bibitem{Keeley:2019esp}
Ryan~E. Keeley, Shahab Joudaki, Manoj Kaplinghat, and David Kirkby.
\newblock {Implications of a transition in the dark energy equation of state
  for the $H_0$ and $\sigma_8$ tensions}.
\newblock 2019.

\bibitem{Gerardi:2019obr}
Francesca Gerardi, Matteo Martinelli, and Alessandra Silvestri.
\newblock {Reconstruction of the Dark Energy equation of state from latest
  data: the impact of theoretical priors}.
\newblock 2019.

\bibitem{Martinelli:2019dau}
Matteo Martinelli, Natalie~B. Hogg, Simone Peirone, Marco Bruni, and David
  Wands.
\newblock {Constraints on the interacting vacuum - geodesic CDM scenario}.
\newblock 2019.

\bibitem{Abbott:2017xzu}
B.~P. Abbott et~al.
\newblock {A gravitational-wave standard siren measurement of the Hubble
  constant}.
\newblock {\em Nature}, 551(7678):85--88, 2017.

\bibitem{TheLIGOScientific:2017qsa}
B.P. Abbott et~al.
\newblock {GW170817: Observation of Gravitational Waves from a Binary Neutron
  Star Inspiral}.
\newblock {\em Phys. Rev. Lett.}, 119(16):161101, 2017.

\bibitem{Monitor:2017mdv}
B.~P. Abbott et~al.
\newblock {Gravitational Waves and Gamma-rays from a Binary Neutron Star
  Merger: GW170817 and GRB 170817A}.
\newblock {\em Astrophys. J.}, 848(2):L13, 2017.

\bibitem{Coulter:2017wya}
D.~A. Coulter et~al.
\newblock {Swope Supernova Survey 2017a (SSS17a), the Optical Counterpart to a
  Gravitational Wave Source}.
\newblock {\em Science}, 2017.

\bibitem{Hotokezaka:2018dfi}
Kenta Hotokezaka, Ehud Nakar, and Ore~\textit{et al.} Gottlieb.

\bibitem{Chen:2017rfc}
Hsin-Yu Chen, Maya Fishbach, and Daniel~E. Holz.
\newblock {A two per cent Hubble constant measurement from standard sirens
  within five years}.
\newblock {\em Nature}, 562(7728):545--547, 2018.

\bibitem{holicowV}
V.~{Bonvin}, F.~{Courbin}, S.~H. {Suyu}, P.~J. {Marshall}, C.~E. {Rusu},
  D.~{Sluse}, M.~{Tewes}, K.~C. {Wong}, T.~{Collett}, C.~D. {Fassnacht},
  T.~{Treu}, M.~W. {Auger}, S.~{Hilbert}, L.~V.~E. {Koopmans}, G.~{Meylan},
  N.~{Rumbaugh}, A.~{Sonnenfeld}, and C.~{Spiniello}.
\newblock {H0LiCOW - V. New COSMOGRAIL time delays of HE 0435-1223: H$_{0}$ to
  3.8 per cent precision from strong lensing in a flat
  {\ensuremath{\Lambda}}CDM model}.
\newblock {\em \mnras}, 465(4):4914--4930, Mar 2017.

\bibitem{Suyu:2016qxx}
S.~H. Suyu et~al.
\newblock {H0LiCOW – I. H0 Lenses in COSMOGRAIL's Wellspring: program
  overview}.
\newblock {\em Mon. Not. Roy. Astron. Soc.}, 468(3):2590--2604, 2017.

\bibitem{Birrer:2018vtm}
S.~Birrer et~al.
\newblock {H0LiCOW - IX. Cosmographic analysis of the doubly imaged quasar SDSS
  1206+4332 and a new measurement of the Hubble constant}.
\newblock {\em Mon. Not. Roy. Astron. Soc.}, 484:4726, 2019.

\bibitem{HolicowXIII}
Kenneth~C. {Wong}, Sherry~H. {Suyu}, and Geoff C. F.~\textit{et a;.} {Chen}.

\bibitem{LSSToverviewpaper}
{\v{Z}}eljko {Ivezi{\'c}}, Steven~M. {Kahn}, and J.~Anthony~\textit{et al.}
  {Tyson}.
\newblock {LSST: From Science Drivers to Reference Design and Anticipated Data
  Products}.
\newblock {\em \apj}, 873(2):111, Mar 2019.

\bibitem{LSST}
Masamune {Oguri} and Philip~J. {Marshall}.
\newblock {Gravitationally lensed quasars and supernovae in future wide-field
  optical imaging surveys}.
\newblock {\em \mnras}, 405(4):2579--2593, Jul 2010.

\bibitem{oguriLSST}
Masamune {Oguri} and Philip~J. {Marshall}.
\newblock {Gravitationally lensed quasars and supernovae in future wide-field
  optical imaging surveys}.
\newblock {\em \mnras}, 405(4):2579--2593, Jul 2010.

\bibitem{liaoLSST}
Kai {Liao}, Tommaso {Treu}, and Phil~\textit{et al.} {Marshall}.
\newblock {Strong Lens Time Delay Challenge. II. Results of TDC1}.
\newblock {\em \apj}, 800(1):11, Feb 2015.

\bibitem{falco1985}
E.~E. {Falco}, M.~V. {Gorenstein}, and I.~I. {Shapiro}.
\newblock {On model-dependent bounds on H(0) from gravitational images
  Application of Q0957 + 561A,B}.
\newblock {\em \apjl}, 289:L1--L4, February 1985.

\bibitem{Grav_lensing1992}
P.~Schneider, J.~Ehlers, and E.E. Falco.
\newblock {\em Gravitational Lenses}.
\newblock Springer, 1992.

\bibitem{Grav_lensing2006}
P.~Schneider, C.~S. Kochanek, and J.~Wambsganss.
\newblock {\em Gravitational Lensing: Strong, Weak and Micro}.
\newblock Springer, 2006.

\bibitem{koopmans2006}
L{\'e}on V.~E. {Koopmans}, Tommaso {Treu}, and Adam S.~\textit{et al.}
  {Bolton}.
\newblock {The Sloan Lens ACS Survey. III. The Structure and Formation of
  Early-Type Galaxies and Their Evolution since z \textasciitilde 1}.
\newblock {\em \apj}, 649(2):599--615, Oct 2006.

\bibitem{Barkana_1998}
Rennan Barkana.
\newblock Fast calculation of a family of elliptical gravitational lens models.
\newblock {\em \apj}, 502(2):531–537, aug 1998.

\bibitem{Saha:2000kn}
P.t Saha, C.~Lobo, A.~Iovino, D.~Lazzati, and G.~Chincarini.
\newblock {Lensing degeneracies revisited}.
\newblock {\em AJ}, 120:1654, 2000.

\bibitem{Wucknitz:2002fd}
O.~Wucknitz.
\newblock {Degeneracies and scaling relations in general power-law models for
  gravitational lenses}.
\newblock {\em \mnras}, 332:951, 2002.

\bibitem{suyu2010}
S.~H. {Suyu}, P.~J. {Marshall}, and M.~W.\textit{et al.} {Auger}.
\newblock {Dissecting the Gravitational lens B1608+656. II. Precision
  Measurements of the Hubble Constant, Spatial Curvature, and the Dark Energy
  Equation of State}.
\newblock {\em \apj}, 711(1):201--221, Mar 2010.

\bibitem{dispersionintegration}
Simon {Birrer}, Adam {Amara}, and Alexandre {Refregier}.
\newblock {The mass-sheet degeneracy and time-delay cosmography: analysis of
  the strong lens RXJ1131-1231}.
\newblock {\em \jcap}, 2016(8):020, Aug 2016.

\bibitem{Keeton:2002ug}
Charles~R. Keeton.
\newblock {Analytic cross-sections for substructure lensing}.
\newblock {\em \apj}, 584:664--674, 2003.

\bibitem{McCully:2013fga}
Curtis McCully, Charles~R. Keeton, Kenneth~C. Wong, and Ann~I. Zabludoff.
\newblock {A New Hybrid Framework to Efficiently Model Lines of Sight to
  Gravitational Lenses}.
\newblock {\em \mnras}, 443(4):3631--3642, 2014.

\bibitem{seljak}
Uros {Seljak}.
\newblock {Large-Scale Structure Effects on the Gravitational Lens Image
  Positions and Time Delay}.
\newblock {\em \apj}, 436:509, Dec 1994.

\bibitem{Suyu:2013kha}
S.~H. Suyu et~al.
\newblock {Cosmology from gravitational lens time delays and Planck data}.
\newblock {\em \apj}, 788:L35, 2014.

\bibitem{galactic_dynamics}
James Binney and Scott Tremaine.
\newblock {\em Galactic Dynamics}.
\newblock Princeton University press, 2008.

\bibitem{Hernquist}
Lars Hernquist.
\newblock {An Analytical Model for Spherical Galaxies and Bulges}.
\newblock {\em \apj}, 356:359, 1990.

\bibitem{Schneider:2013sxa}
Peter Schneider and Dominique Sluse.
\newblock {Mass-sheet degeneracy, power-law models and external convergence:
  Impact on the determination of the Hubble constant from gravitational
  lensing}.
\newblock {\em \aap}, 559:A37, 2013.

\bibitem{Hu:2013twa}
Bin Hu, Marco Raveri, Noemi Frusciante, and Alessandra Silvestri.
\newblock {Effective Field Theory of Cosmic Acceleration: an implementation in
  CAMB}.
\newblock {\em Phys. Rev.}, D89(10):103530, 2014.

\bibitem{Raveri:2014cka}
Marco Raveri, Bin Hu, Noemi Frusciante, and Alessandra Silvestri.
\newblock {Effective Field Theory of Cosmic Acceleration: constraining dark
  energy with CMB data}.
\newblock {\em Phys. Rev.}, D90(4):043513, 2014.

\bibitem{CAMB1}
Antony Lewis, Anthony Challinor, and Anthony Lasenby.
\newblock {Efficient computation of CMB anisotropies in closed FRW models}.
\newblock {\em Astrophys. J.}, 538:473--476, 2000.

\bibitem{CAMB2}
Cullan Howlett, Antony Lewis, Alex Hall, and Anthony Challinor.
\newblock {CMB power spectrum parameter degeneracies in the era of precision
  cosmology}.
\newblock {\em \textit{J. Cosmology Astropart. Phys.}}, 1204:027, 2012.

\bibitem{cosmomc}
Antony Lewis and Sarah Bridle.
\newblock {Cosmological parameters from CMB and other data: A Monte Carlo
  approach}.
\newblock {\em \prd}, 66:103511, 2002.

\bibitem{cp}
Michel {Chevallier} and David {Polarski}.
\newblock {Accelerating Universes with Scaling Dark Matter}.
\newblock {\em International Journal of Modern Physics D}, 10(2):213--223, Jan
  2001.

\bibitem{l}
Eric~V. Linder.
\newblock Exploring the expansion history of the universe.
\newblock {\em \prl}, 90:091301, Mar 2003.

\bibitem{holicowIV}
Kenneth~C. {Wong}, Sherry~H. {Suyu}, and Matthew W.~\textit{et al.} {Auger}.
\newblock {H0LiCOW - IV. Lens mass model of HE 0435-1223 and blind measurement
  of its time-delay distance for cosmology}.
\newblock {\em \mnras}, 465(4):4895--4913, Mar 2017.

\bibitem{Oguri:2005je}
Masamune Oguri.
\newblock {The image separtion distribution of strong lenses: halo versus
  subhalo populations}.
\newblock {\em Mon. Not. Roy. Astron. Soc.}, 367:1241--1250, 2006.

\bibitem{quasarredshift}
Isabelle {P{\^a}ris}, Patrick {Petitjean}, and {\'E}ric~\textit{et al.}
  {Aubourg}.
\newblock {The Sloan Digital Sky Survey Quasar Catalog: Fourteenth data
  release}.
\newblock {\em \aap}, 613:A51, May 2018.

\bibitem{Rusu:2006}
Cristian~E. {Rusu}, Christopher~D. {Fassnacht}, and Dominique~\textit{et al.}
  {Sluse}.
\newblock {H0LiCOW - III. Quantifying the effect of mass along the line of
  sight to the gravitational lens HE 0435-1223 through weighted galaxy
  counts★}.
\newblock {\em \mnras}, 467(4):4220--4242, Jun 2017.

\bibitem{plank2018}
N.~\textit{et al.} Aghanim.
\newblock {Planck 2018 results. VI. Cosmological parameters}.
\newblock 2018.

\bibitem{hojjati}
Alireza {Hojjati} and Eric~V. {Linder}.
\newblock {Next generation strong lensing time delay estimation with Gaussian
  processes}.
\newblock {\em \prd}, 90(12):123501, Dec 2014.

\bibitem{Jee_2016}
I.~Jee, E.~Komatsu, S.H. Suyu, and D.~Huterer.
\newblock Time-delay cosmography: increased leverage with angular diameter
  distances.
\newblock {\em \jcap}, 2016(04):031--031, apr 2016.

\bibitem{Albrecht:2006}
Andreas {Albrecht}, Gary {Bernstein}, and Robert~\textit{et al.} {Cahn}.
\newblock {Report of the Dark Energy Task Force}.
\newblock {\em arXiv e-prints}, pages astro--ph/0609591, Sep 2006.

\bibitem{2011arXiv1110.3193L}
R.~{Laureijs}, J.~{Amiaux}, and S.~\textit{et al.} {Arduini}.
\newblock {Euclid Definition Study Report}.
\newblock {\em arXiv e-prints}, page arXiv:1110.3193, Oct 2011.

\bibitem{Bacon:2018dui}
David~J. Bacon et~al.
\newblock {Cosmology with Phase 1 of the Square Kilometre Array: Red Book 2018:
  Technical specifications and performance forecasts}.
\newblock {\em Submitted to: Publ. Astron. Soc. Austral.}, 2018.

\bibitem{MGtest1}
Thomas~E. {Collett}, Lindsay~J. {Oldham}, and Russell J.~\textit{et al.}
  {Smith}.
\newblock {A precise extragalactic test of General Relativity}.
\newblock {\em Science}, 360(6395):1342--1346, Jun 2018.

\bibitem{MGtest2}
Tristan~L. {Smith}.
\newblock {Testing gravity on kiloparsec scales with strong gravitational
  lenses}.
\newblock {\em arXiv e-prints}, page arXiv:0907.4829, Jul 2009.

\bibitem{MGtest3}
Dhrubo {Jyoti}, Julian~B. {Mu{\~n}oz}, Robert~R. {Caldwell}, and Marc
  {Kamionkowski}.
\newblock {Cosmic time slip: Testing gravity on supergalactic scales with
  strong-lensing time delays}.
\newblock {\em \prd}, 100(4):043031, Aug 2019.

\end{thebibliography}

\end{document}